\documentclass[aps,twocolumn,longbibliography, superscriptaddress,english,prx]{revtex4}
\usepackage[colorlinks=true,urlcolor=blue,citecolor=blue,linkcolor=blue]{hyperref}
\usepackage[T1]{fontenc}
\usepackage{amssymb}
\usepackage{graphicx}
\usepackage{subfigure}
\usepackage{pifont}
\usepackage{amsmath,color, bm}
\usepackage{mathrsfs}
\usepackage{float}
\usepackage[normalem]{ulem}
\usepackage{indentfirst}
\usepackage{txfonts}
\usepackage{algpseudocode}
\usepackage{algorithm}
\usepackage{booktabs}

\emergencystretch=\maxdimen
\makeatletter

\def\journal #1, #2, #3, 1#4#5#6{{\sl #1~}{\bf #2}, #3 (1#4#5#6) }

\newcommand{\material}[1]{\iffalse[{\bf  \color{cyan}{Material: #1}}]\fi}

\usepackage{amsmath, amssymb}

\makeatother

\usepackage{babel}
\begin{document}
\title{Simulating noisy variational quantum eigensolver with local noise models}

\author{Jinfeng Zeng}
\affiliation{Department of Physics, The Hong Kong University of Science and Technology, Clear Water Bay, Kowloon, Hong Kong}

\author{Zipeng Wu}
\affiliation{Department of Physics, The Hong Kong University of Science and Technology, Clear Water Bay, Kowloon, Hong Kong}

\author{Chenfeng Cao}
\affiliation{Department of Physics, The Hong Kong University of Science and Technology, Clear Water Bay, Kowloon, Hong Kong}

\author{Chao Zhang}
\affiliation{Center for Artificial Intelligence, Peng Cheng Laboratory, Shenzhen, 518055, China}

\author{Shi-Yao Hou}
\affiliation{Department of Physics, The Hong Kong University of Science and Technology, Clear Water Bay, Kowloon, Hong Kong}
\affiliation{College of Physics and Electronic Engineering, Center for Computational Sciences,  Sichuan Normal University, Chengdu 610068, China}

\author{Pengxiang Xu}
\affiliation{Center for Artificial Intelligence, Peng Cheng Laboratory, Shenzhen, 518055, China}

\author{Bei Zeng}
\email[]{zengb@ust.hk}
\affiliation{Department of Physics, The Hong Kong University of Science and Technology, Clear Water Bay, Kowloon, Hong Kong}

\begin{abstract}
The variational quantum eigensolver (VQE) is a promising algorithm to demonstrate quantum advantage on near-term noisy-intermediate-scale quantum (NISQ) computers. One central problem of VQE is the effect of noise, especially the physical noise, on realistic quantum computers. We systematically study the effect of noise for the VQE algorithm by performing numerical simulations with various local noise models, including the amplitude damping, dephasing, and depolarizing noise.  We show that the ground state energy will deviate from the exact value as the noise probability increase, and typically, the noise will accumulate as the circuit depth increase. The results suggest that the noisy quantum system can remain entanglement at the noise level of NISQ devices by comparing the VQE solution with the mean-field solution for the many-body ground state problem. We build a noise model to capture the noise in a real quantum computer, and the corresponding numerical simulation is consistent with experimental results on IBM Quantum computers through Cloud. Our work sheds new light on the practical research of noisy VQE, and the deep understanding of the noise effect of VQE will also help with developing error mitigation techniques on near-term quantum computers.
\end{abstract}

\maketitle

\section{Introduction}
Quantum computers are believed to be the next generation of computing devices. There are efficient quantum algorithms that can achieve an exponential speedup compared with classical algorithms, such as Shor's algorithm for large number factoring~\cite{shor1999polynomial}, and the HHL algorithm for solving linear systems of equations~\cite{harrow2009quantum}. In practice, quantum systems are fragile when interacting with the environment, which inevitably causes decoherence. Theory on quantum error correction and fault-tolerance has been developed \cite{Shor1995Scheme, Steane1996Error, Gottesman1996Class, Bennett1996Mixed, Knill1997Theory, gaitan2008quantum}, which guarantees that under reasonable assumptions of the noise models, arbitrarily long quantum computing could be reliable provided that the error per gate is less than some threshold value. However, the overhead to achieve fault tolerance is significant, and it is not expected to be within reach of near-future technology.

At the moment, we are entering the so-called noisy-intermediate-scale quantum (NISQ) machines era~\cite{preskill2018quantum}.  A fundamental challenge is to show quantum advantage on near-term NISQ devices. Recently, a superconducting quantum processor \cite{Google2019Supremacy} realizes random quantum circuit sampling, and a photonic quantum computer \cite{Zhong2020Advantage} performed boson sampling to show a quantum advantage. These specifically designed tasks can achieve exponential speedup on quantum computers comparing to the classical algorithm. It is important to develop quantum algorithms suitable for NISQ devices to solve real-world applications. Many hybrid quantum-classical algorithms are developed to adapt to near-term noisy devices, such as the variational quantum eigensolver (VQE) \cite{Peruzzo2014Variational, McClean2016VQE, Kandala2017},
quantum approximate optimization algorithm (QAOA) \cite{farhi2014quantum}, quantum imaginary time evolution (QITE) \cite{motta2020determining, McArdle2019QITE}, and variational quantum Boltzmann machine~\cite{zoufal2020variational,shingu2020boltzmann}.
One central problem regarding these hybrid algorithms is the effect of noise. Whether the quantum feature, such as entanglement, can be maintained in the noisy quantum circuit. It is crucial to show quantum entanglement in the NISQ devices~\cite{Wang2018IBM, mooney2021wholedevice}. 

There are some arguments that the quantum-classical algorithms are robust to certain types of errors. Ref.~ \cite{xue2019effects, Marshall2020Characterizing, wang2020noise} studied the objective function's flattening due to noise in the variational quantum algorithm.
 In Ref.~\cite{sharma2020noise}, the authors concluded that variational quantum compiling algorithms are resilient to noise.  Ref.~\cite{gentini2019noiseassisted} studied the optimization process of variational hybrid quantum-classical algorithms from the theoretical viewpoint and found that the noise can be beneficial to some regimes' optimization problems.  VQE is a promising variational quantum algorithm with a broad range of applications~\cite{cerezo2020variational, bravo2019variational, wang2020variational, anschuetz2019variational, zeng2020variational, larose2019variational, bravo2020quantum}. In this paper, we focus on the effect of noise on the VQE algorithm and explore the quantumness of VQE with noise.  Some experiments use the VQE algorithm to calculate electronic structures \cite{Kandala2017, o2016scalable, Hempel2018Quantum, arute2020hartree}, demonstrating the algorithm's robustness against noise to some extend.  Ref.~\cite{McClean2016VQE}  theoretically argued that the VQE could suppress the errors by utilizing the ansatz's symmetry. However, it did not study the physical error effect in the realistic quantum device. Ref.~\cite{Kandala2017} numerically study the optimization versus decoherence effects' behavior, yet the effect of realistic noise for VQE, in general, remains unclear.   Moreover, it is hard to simulate quantum circuits on classical computers in both the exact and noisy regime \cite{Bremner2011Classical,aaronson2011computational, Fujii2017Commuting, Bremner2016Average,aaronson2016complexitytheoretic}.


In this work, we systematically study the noise effect of the VQE algorithm and investigate the quantum entanglement of the noisy quantum system by performing simulations with various local noise models. These local noise models include amplitude damping, dephasing, and depolarizing noise existing in realistic quantum devices. We numerically simulate the noisy VQE to solve the ground state energy problem of different lattice spin models with finite correlation length. We found that the noise will gradually deviate from the ground state energy as the noise probability increases. Besides, the noise will accumulate as the circuit depth increase.  We then compare the noisy VQE solution with the classical mean-field solution, which means variational space is a separable state without quantum entanglement. If the noisy VQE can outperform the mean-field solution, this indicates the quantumness, such as quantum entanglement, of the noisy system be maintained. Our results suggest that the noisy quantum system can remain entanglement at near-term quantum computers' noise level. We also build a noise model to capture the IBM quantum computer's realistic noise and compare the numerical results with the experimental results on IBM Quantum Experience through Cloud. Our numerical simulations are consistent with the quantum experiment results. Our work sheds new light on the practical research of noisy VQE. A deep understanding of the effect of noise for VQE may help develop quantum error mitigation techniques \cite{kandala2019error, Temme2017Error, Endo2018Practical}  on near team quantum devices.

\section{Methods}
In this section, We briefly introduce the VQE algorithm and the noise model we use in this work.

\subsection{The VQE method}
\begin{figure}[tbh]
\centerline{\includegraphics[width=0.50\textwidth]{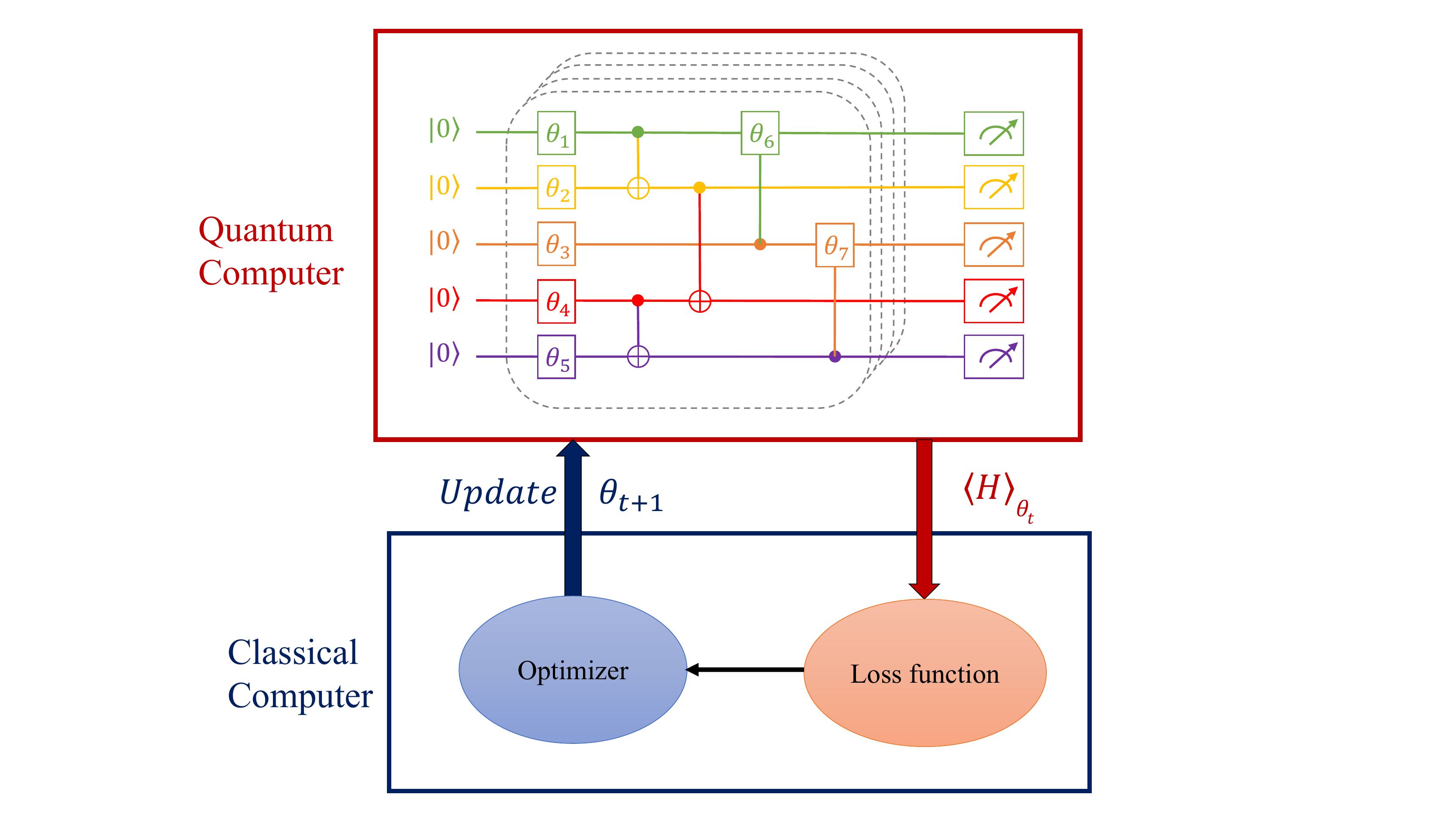}}
\caption{Schematic illustration of a quantum-classical hybrid algorithm. At step $t$, the quantum circuit is controlled by the classical parameters $\bm{\theta}_{t}$. The parameters and the structure of the quantum circuit define the variational ansatz. Measurement output of the quantum circuit then defines a loss function according to the problem (for example, the energy of a given Hamiltonian $\langle H \rangle_{\bm{\theta}_{t}}$). Optimize the loss function with respect to the parameters on a classical computer and update the parameters of the quantum circuit as $\bm{\theta}_{t+1}$. Repeat the above procedures until the condition of convergence.}
\label{QNN-fig}
\end{figure}

The VQE~\cite{Peruzzo2014Variational,McClean2016VQE,Kandala2017, cerezo2020variational, bravo2019variational, wang2020variational, anschuetz2019variational, zeng2020variational, larose2019variational, bravo2020quantum}  is a quantum-classical hybrid algorithm to solve the quantum eigen problem, which can be run on NISQ devices.  The quantum-classical hybrid algorithm runs on both quantum computer and classical computer. We first construct a quantum circuit, which is controlled by the classical parameters.  The quantum circuit output can be changed by modifying the classical parameters. According to the problem, we define a loss function with respect to the parameters and then use a classical optimizer to update the parameters. After calculating the new parameters by the classical computer, we update the parameters in the quantum circuit. Repeat the above procedures until the condition of convergence. Fig.~\ref{QNN-fig}  illustrates the main procedure of the quantum-classical hybrid algorithm.

The major components of the quantum-classical hybrid algorithm include the quantum circuit ansatz, the loss function, and the optimizer. When we consider the ground state energy problem with VQE, the loss function is chosen as
\begin{eqnarray}
E(\bm{\theta}) &=& \langle 0 | U^{\dagger}(\bm{\theta})H U(\bm{\theta})| 0 \rangle = \langle  H  \rangle_{\bm{\theta}}, 
\label{Venergy}
\end{eqnarray}
where $\bm{\theta}$ are the parameters of the quantum circuit, $U(\bm{\theta})$ is the unitary transformation. The classical optimizer can use a gradient-free method or gradient method.  Since one can not directly access the wave function on a real quantum device, we have neither the analytical form of the loss function nor the loss function's gradient. However, we can also use the gradient-based optimizer  by making use of the formula~\cite{Schuld2019,Mitarai2018}
\begin{eqnarray}
\frac{\partial E(\bm{\theta})}{\partial \theta_j} &=& (\langle  H  \rangle_{\bm{\theta}_j^{+}} - \langle  H  \rangle_{\bm{\theta}_j^{-}})/2, 
\label{gradient}
\end{eqnarray}
where $\bm{\theta}_j^{\pm} = \bm{\theta} \pm \frac{\pi}{2}\bm{e}_{j}$, with $\bm{e}_{j}$ being the $j$-th unit vector in parameter space. The unbaised gradient estimator ( Eq.~\eqref{gradient} ) is necessary, because the noise error and the sampling error will make the finite difference gradient estimator have poor performance in the real quantum experiment.
We can access the analytical form of the wave function as well as the variational energy ( Eq.~\eqref{Venergy} ) in simulation. So we can use all of the classical optimization methods such as gradient descent (GD), Adam~\cite{Kingma2014Adam}, and Broyden-Fletcher-Goldfarb-Shanno (BFGS) method in simulation.

It is challenging to choose the variational ansatz $U(\bm{\theta})$ for general Hamiltonian~\cite{Cao2019Quantum,Sim2019Expressibility}. Two different types of quantum circuit ansatz, physically motivated ansatz, and hardware efficient ansatz, have been proposed according to different considerations. The physically motivated ansatz, including unitary coupled-cluster~\cite{McClean2016VQE, Romero2018Strategies}, Hamiltonian variational ansatz~\cite{Wecker2015Progress}, and low-depth circuit ansatz~\cite{Dallaire2019LDCA},  are based on or inspired by the original problem that systematically approximates the exact electronic wave function. The hardware efficient ansatz~\cite{Kandala2017} comprises a series of parametrized single-qubit gates and two-qubit gates, which can be directly implemented on the existing NISQ hardware. Besides the above fixed ansatz approaches, one can adopt the Adaptive Derivative-Assembled Pseudo-Trotter VQE (ADAPT-VQE) methods~\cite{Grimsley2019Adaptive,tang2020qubitadaptvqe}, in which the ansatz is dynamically built.
In this work, we consider to simulate and analyze the noisy VQE of one-dimensional spin models. We also run the VQE algorithm on a real IBM quantum computer to compare with our simulation, so it is suitable to use the hardware efficient ansatz. 

\subsection{Noise model}\label{sec:noise_model}
In this section, we describe the gate noise models for circuit simulation. For any state $\rho$ of the system, the noise channel $\varepsilon$  can be described by the Kraus representation~\cite{nielsen_chuang_2010}
\begin{eqnarray}
\varepsilon(\rho) =& \sum_{k} E_k \rho E_k^{\dagger},
\end{eqnarray}
where $E_{k}$s are the Kraus operators and $\sum_{k} E_k E_k^{\dagger} = I$. 

The hardware efficient quantum circuits usually contain single-qubit and two-qubit quantum gates. In this paper, We model the noise only on single-qubit and two-qubit quantum gates by 
\begin{eqnarray}\label{noise_model}
\rho \rightarrow  \sum_{k} E_k U \rho U^{\dagger} E_k^{\dagger} = \sum_{k} E_k  \rho' E_k^{\dagger} ,
\end{eqnarray}
where $E_k$s acting on the same single qubit and two qubits as $U$ acts on. In this paper, We consider the following three types of noise models.

\begin{itemize}
\item Amplitude damping noise

The amplitude damping noise~\cite{nielsen_chuang_2010} can be characterized by the Kraus operators,
\begin{displaymath}
K_{1}=\left(
\begin{array}{cc}
1 & 0    \\
0  & \sqrt{1-p}   \\
\end{array}
\right), 
K_{2}=\left(
\begin{array}{cc}
0 & \sqrt{p}    \\
0  & 0  \\
\end{array}
\right),
\end{displaymath}

where $p \in [0, 1]$ is the probability of the noise.
For amplitude damping noise on single-qubit gate $U$, the Kraus operators $E_k$s in Eq.~\eqref{noise_model} run over the set $\{K_1, K_2\}$.
The Kraus operators $E_k$s run over the set $\{K_1, K_2\}\otimes\{K_1, K_2\}$ for two-qubit noisy gate, which is equal to,
\begin{eqnarray}
\{K_1 \otimes K_1, K_1 \otimes K_2, K_2 \otimes K_1, K_2 \otimes K_2\}. \nonumber
\end{eqnarray}

\item Dephasing noise

The  dephasing noise~\cite{nielsen_chuang_2010}  is characterized by the Kraus operators,
\begin{eqnarray}
K_{1}=\sqrt{1-p}I_{2}, \quad  K_{2}=\sqrt{p}\sigma_{Z},
\end{eqnarray}
where $I_{2}$ is the two dimensional identity matrix and $\sigma_{Z}$ is Pauli operator. For dephasing  noise on single-qubit gate $U$, the Kraus operators $E_k$s in Eq.~\eqref{noise_model} run over the set $\{K_1, K_2\}$. The Kraus operators $E_k$s run over the set $\{K_1, K_2\}\otimes\{K_1, K_2\}$ for two-qubit noisy gate.
  
\item Depolarizing noise

The depolarizing noise ~\cite{nielsen_chuang_2010} on a single qubit gate is modeled by
\begin{eqnarray}
\rho &\rightarrow &  (1-p) U \rho U^{\dagger} +  p \frac{I}{2}  \nonumber \\
&=& (1-\frac{3}{4}p)\rho' + \frac{p}{4}(X \rho' X^{\dagger} + Y \rho' Y^{\dagger} + Z \rho' Z^{\dagger}),
\end{eqnarray}
where $\rho' = U \rho U^{\dagger} $ and $X,Y,Z$ are Pauli operators. 
For depolarizing noise on single-qubit gate $U$, the Kraus operators $E_k$s in Eq.~\eqref{noise_model} run over the set 
\begin{equation}
\{K_1 = \sqrt{1-3p/4}I, K_2 = \frac{\sqrt{p}}{2}X, K_3 = \frac{\sqrt{p}}{2}Y,  K_4 = \frac{\sqrt{p}}{2}Z \}\nonumber
\end{equation}
The Kraus operators $E_k$s run over the set $\{K_1, K_2, K_3, K_4\}\otimes\{K_1, K_2, K_3, K_4\}$ for two-qubit noisy gate.
\end{itemize}

\section{Results}
\begin{figure}[tb]
\centerline{\includegraphics[width=0.49\textwidth]{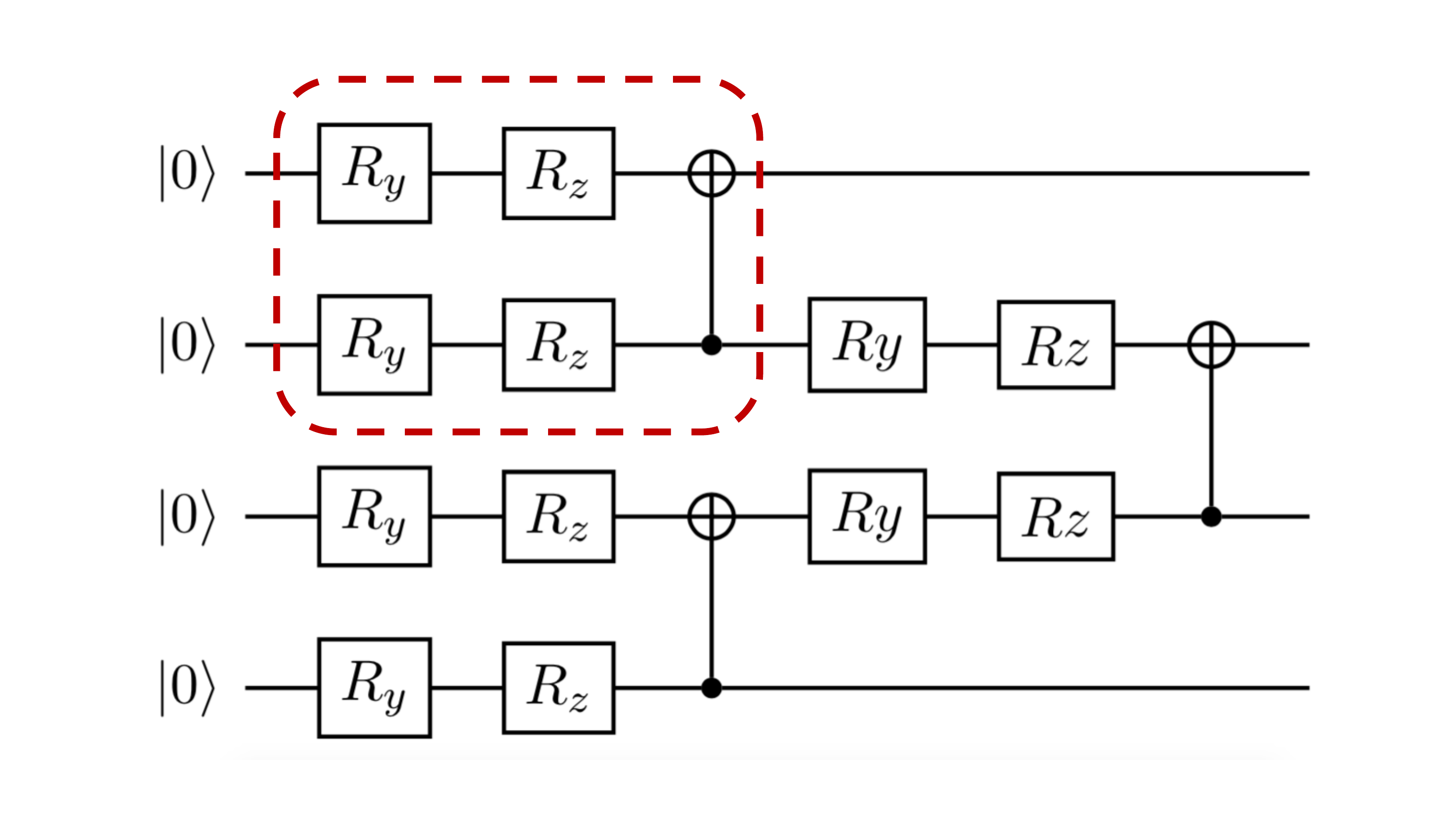}}
\caption{The sketch of the quantum circuit ansatz used for the noisy VQE simulation. This is a $4$-qubit quantum circuit with one ``logical" circuit depth. This one logical depth of the $2n$-qubit quantum circuit includes two physical layers. The first layer is comprised of $n$ blocks. Each block is indicated by the red dotted frame. The $i$-th block is comprised of  CNOT on the $(2i, 2i+1)$ qubits following $2$ parameterized single-qubit rotation gates $R_{y}$ and $R_{z}$ on each qubit, where $i \in [0, n-1]$ . The second physical layer is comprised of $n-1$ blocks. The $j$-th block in second layer act on $(2j+1, 2j+2)$ qubits, where $j \in [0, n-2]$. Stack $d$ logical layer to compose the whole quantum circuit ansatz. The $2n$-qubit  ansatz with $d$ circuit depth hence has $(2n-1)\times 4d$ learnable parameters. }
\label{ansatz}
\end{figure}

We simulate the noisy VQE on one-dimensional models and analyze the noise effect of the VQE algorithm. We consider three one-dimensional spin models, including the transverse field Ising model, Heisenberg model, and the transverse field Heisenberg model. We choose the heuristic hardware-efficient ansatz shown in Fig.~\ref{ansatz},  which has the suitable expressibility and entangling capability~\cite{Sim2019Expressibility} for the one-dimensional spin models.

\begin{figure*}[tb]
\centerline{\includegraphics[width=0.98\textwidth]{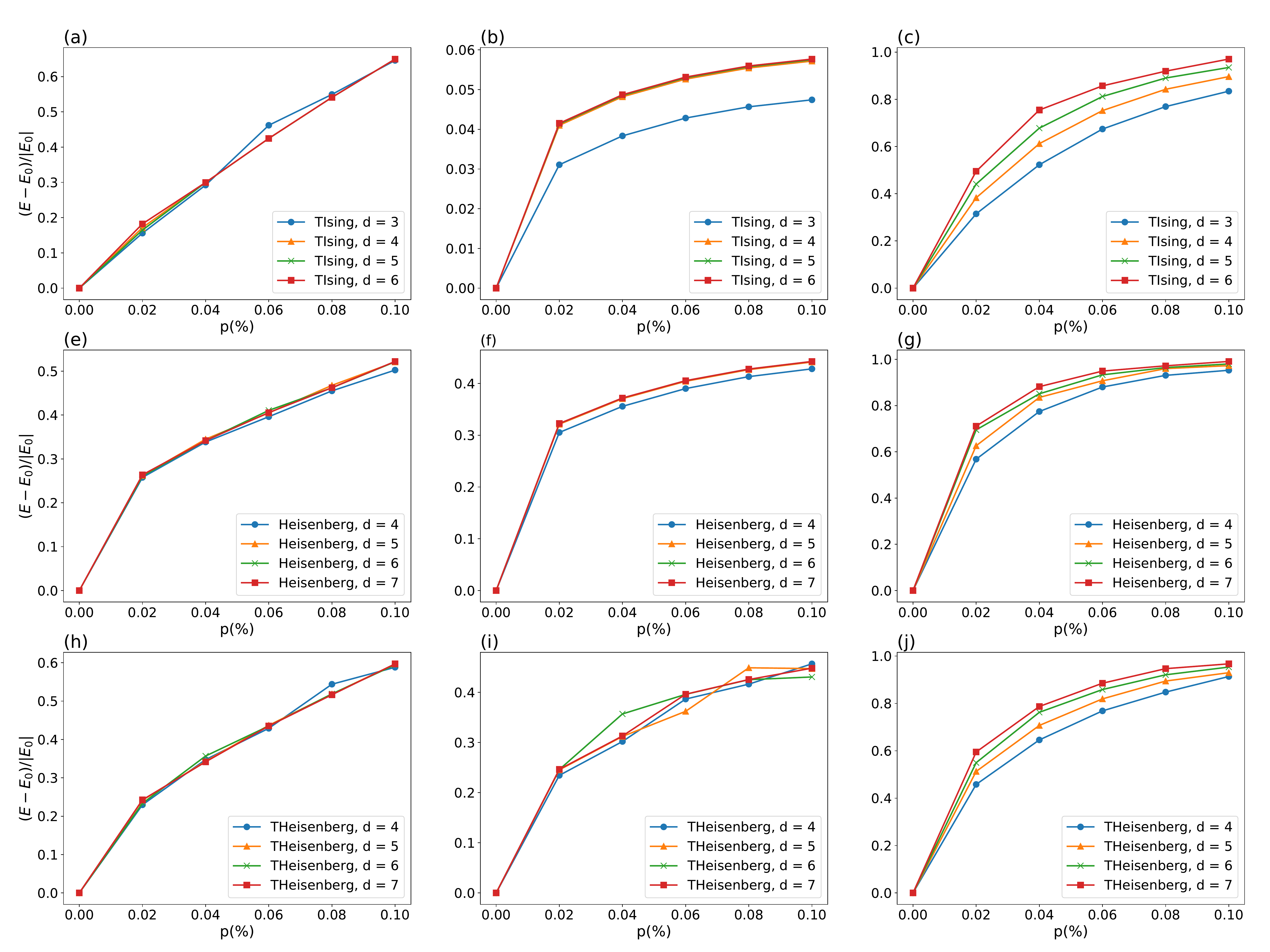}}
\caption{The numerical results of noisy VQE simulation for the transverse field Ising model, Heisenberg model and transverser field Heisenberg model with number of qubit $n=6$. (a) The amplitude damping noise, (b) the dephasing noise  and (c) the depolarizing noise apply in the quantum circuit for  transverse field Ising model with $J=h=1$ in Eq.~\ref{TIsing}. (e) The amplitude damping noise, (f) the dephasing noise and (g) the depolarizing noise apply in the quantum circuit for  transverse field Ising model with $J=1, h=0$ in Eq.~\ref{THeisenberg}. (h) The amplitude damping noise, (i) the dephasing noise  and (j) the depolarizing noise apply in the quantum circuit for  transverse field Heisenberg model with $J=h=1$ in Eq.~\ref{THeisenberg}. $p$ is noise probability with unit $\%$. The y axis is the relative energy between energy $E$ given by the noise VQE and the exact energy, $(E-E_{0})/|E_{0}|$.}
\label{fig:noise_effect}
\end{figure*}

The transverse field Ising model is defined as
\begin{eqnarray}
\label{TIsing}
H = J\sum_{j}^{N} \sigma_{j}^{x}  \sigma_{j+1}^{x}  + h \sum_{j}^{N} \sigma_{j}^{z}.
\end{eqnarray}
The Heisenberg model is defined as
\begin{eqnarray}
\label{THeisenberg}
H = \sum_{j}^{N} \boldsymbol{S}_{i} \cdot  \boldsymbol{S}_{j}.
\end{eqnarray}
The transverse field Heisenberg model is defined as
\begin{eqnarray}
\label{Heisenberg}
H = J\sum_{j}^{N} \boldsymbol{S}_{i} \cdot  \boldsymbol{S}_{j} + h \sum_{j}^{N} \sigma_{j}^{z}.
\end{eqnarray}
In the numerical experiment of transverse field Ising model and transverse field Heisenberg, we set $J=h=1$. When $N>2$, we apply the periodic boundary condition.

We firstly run the VQE algorithm without noise to get energy with a different number of qubits $n$.  In each $n$,  we increase the circuit depth and find the minimal logical circuit depth in which the energy of VQE is approximate to the exact energy within $98\%$, meaning that the ratio between energy $E$ given by the VQE and the exact ground state energy $E_0$, $E/E_{0} \geq 98\%$.  The results are shown in Table~\ref{tab:circuit-depth}.

\begin{table}[h!]
\centering
\addtolength{\tabcolsep}{18pt}
\caption{ The minimal logical circuit depth $d$ needed for the VQE energy approximate the exact energy within $98\%$ for $n$ qubits ( ratio between energy $E$ given by the VQE and the exact ground state energy $E_0$, $E/E_{0} \geq 98\%$). $n$ is the number of qubits.}
\label{tab:circuit-depth}
\begin{tabular}{@{}lllllll@{}}
\toprule
 n   & 2  & 4   & 6 &  8  \\ \midrule
TIsing       & 1  & 2   & 3 & 3  \\ \midrule
Heisenberg   & 1  & 3   & 4 & 6  \\ \midrule
THeisenberg   & 1  & 3   & 4 & 6 \\ \bottomrule
\end{tabular}
\end{table}



We then apply the amplitude damping, dephasing, and depolarizing noise model in the quantum circuit as described in Sec.~\ref{sec:noise_model}. In each $n$-qubit experiment, we start with the minimal logical circuit depth $n$ and add the circuit depth. Each data point is obtained by averaging over 3 optimizations with different random initial parameters. Fig.~\ref{fig:noise_effect} shows the numerical results of noisy VQE simulations for the transverse field Ising model, the Heisenberg model, and the transverse field Heisenberg model. To characterize the noise effect, we define $(E-E_{0})/|E_{0}|$ to see how far the energy $E$ given by noisy VQE simulation is different from the exact energy $E_{0}$.  As shown in Fig.~\ref{fig:noise_effect}, as the noise probability increase, the noisy VQE energy is getting further and further away from the exact value and the $(E-E_{0})/|E_{0}|$ increase. Note that different noise model has a different impact on the VQE. Comparing Fig.~\ref{fig:noise_effect} (a), (b) with (c),  the relative energy $(E-E_{0})/|E_{0}|$ in depolarizing noise is larger than the one in damping and dephasing noise for transverse field Ising model, which clearly shows an effect of noise accumulate over depth as one will typically expect. The Heisenberg and the transverse field Heisenberg model share a similar phenomenon with less noise accumulation over depth. It means that the VQE is more sensitive to depolarizing noise than damping and dephasing noise (see an explanation in Appendix A).

\section{explore the quantumness of noisy circuit}
In this section, we explore the quantum feature of noisy VQE.  Entanglement is an important feature that the quantum computers can solve problems intractable for classical computers. It is interesting to ask whether the noisy quantum circuit can maintain quantum entanglement and how much noise will destroy the entanglement. We compare the noisy VQE and classical mean-field solution for ground state energy of spin model with finite correlation length. If the energy obtained by noisy VQE is lower than the classical mean-field energy, the quantum entanglement is still maintained. The results suggest that the VQE can outperform the mean-field solution at near-term quantum computers' noise level.

The single-qubit wave function can be parameterized as
\begin{eqnarray}
|\psi \rangle  = \cos \frac{\theta}{2} |0\rangle + e^{i \varphi}  \sin \frac{\theta}{2} | 1 \rangle.
\end{eqnarray}
The mean-field method to solve the many-particle problem, such as the many-particle ground state energy problem, is to ignore the correlation between particles and assume the many-body wave function as this formula
\begin{eqnarray}
| \Psi \rangle = |\psi_1 \rangle \otimes  |\psi_2 \rangle \otimes ... \otimes  |\psi_n \rangle.
\end{eqnarray}
Given a k-local Hamiltonian $H=\sum_{i} H_i$, where each term have the form   $H_i = \sum_{\alpha\beta}\lambda_{\alpha\beta} \sigma^{\alpha}_{i}\sigma^{\beta}_{j}$  for  2-local Hamiltonian. The mean field energy is defined as 
\begin{eqnarray}
 E_{\text{mf}} &=& \langle \Psi | H | \Psi \rangle  =  \sum_{i}  \langle \Psi | H_i | \Psi \rangle, \\ 
 \langle \Psi | H_i | \Psi \rangle  &=& \sum_{\alpha\beta} \lambda_{\alpha\beta}  \langle \Psi | \sigma^{\alpha}_i \sigma^{\beta}_j | \Psi \rangle  \nonumber \\
&=& \sum_{\alpha\beta} \lambda_{\alpha\beta}  \langle \Psi_{i} | \sigma^{\alpha}_i  | \Psi_{i} \rangle \otimes  \langle \Psi_{j} | \sigma^{\beta}_j  | \Psi_{j} \rangle.
\end{eqnarray}
The mean-field many-body energy is then decoupled to the sum of each term and we just need to calculate the energy of each term and the gradient of the energy with respect to the parameter $(\theta_{i}, \varphi_{i})$ and $(\theta_{j}, \varphi_{j})$. To be specific, each term have the analytical form
\begin{eqnarray}
&&L(\theta_i, \varphi_i) = \langle \Psi_{i} | \sigma^{\alpha}_i  | \Psi_{i} \rangle  = \cos^2 \frac{\theta_i}{2}\langle 0 |\sigma^{\alpha}_{i} |0 \rangle  \nonumber \\
 && + \sin^{2} \frac{\theta_i}{2}\langle 1 |\sigma^{\alpha}_{i} | 1\rangle + \cos \varphi_i \sin \theta_i \langle 0 |\sigma^{\alpha}_{i} | 1\rangle,
\end{eqnarray}
where $\sigma^{\alpha}_{i}$ is one of the Pauli matrix on $i$-th particle. The gradients are
\begin{eqnarray}
\frac{\partial L(\theta, \varphi)}{\partial \theta} &=&  \frac{ \sin \theta}{2} \left( \langle 1 |\sigma | 1\rangle - \langle 0 |\sigma | 0\rangle \right) \nonumber \\
&&+ \cos \varphi \cos \theta  \langle 0 |\sigma | 1 \rangle , \\
\frac{\partial L(\theta, \varphi)}{\partial \varphi} &=& -\sin \varphi \sin \theta  \langle 0 |\sigma | 1 \rangle.
\end{eqnarray}
It is easy to calculate the mean-field ground state energy with the help of the above equations.


\begin{table}[h!]
\centering
\addtolength{\tabcolsep}{12pt}
\caption{The probability of depolarizing noise that allows the noisy VQE can surpass the mean-field solution for the transverse field Ising model,  the Heisenberg model, and the transverse field Heisenberg model.}
\label{tab:surpass_mf}
\begin{tabular}{@{}lllllll@{}}
\toprule
nspin              & 2         & 4         & 6         & 8\\ \midrule
\textbf{TIsing} \\
depolarizing  &0.024   &0.041      &0.027   &0.025    \\ \midrule
\textbf{Heisenberg} \\
depolarizing  & 0.205   & 0.022     &0.011    & 0.005   \\ \midrule
\textbf{THeisenberg} \\
depolarizing  & 0.206 &  0.034   &  0.025   &  0.014   \\ \bottomrule
\end{tabular}
\end{table}

To compare with the VQE method on a quantum computer with noise and the classical mean-field method. We simulate the noisy VQE with the circuit depth given in Tabel~\ref{tab:circuit-depth} on the one-dimensional transverse field Ising model,  Heisenberg model with and without transverse field. We fit the simulated data to the function $y = ce^{ax+b}$, where $y$ is the energy and $x$ is the probability of the noise model.  Then, we have the mean-field energy and the noise VQE energy as a function of the noise model's probability.  Finally, we solve the noise probability that makes the VQE solution surpass the mean-field solution, which means if the noise probability is below the value given in Table~\ref{tab:surpass_mf},  then the noisy VQE energy is lower than the mean-field energy.

As discuss above, the damping and dephasing noises have a more negligible effect on VQE than depolarizing. In this section, we analyze the quantumness of VQE with depolarizing noise. The probability value of depolarizing decreases as the number of qubit increases. When the number of qubit $n=8$, the probability that makes noisy VQE surpass mean-field is around $0.5\%-2.5\%$, which is close to the noise parameters of the near term NISQ devices.

\section{Comparison  with experiment}

\subsection{Experimental setup}

We perform the VQE experiments on realistic quantum computers through IBM Quantum Experience  (IBMQ) \cite{ibmq}, a quantum cloud service released by IBM. It provides several quantum processors with different qubit sizes and noise levels. Many quantum experiments have been carried out on IBMQ.

As mentioned above, we aim at solving the optimization problem 
$\boldsymbol{\theta}_{\mathrm{opt}}=\arg \min _{\boldsymbol{\theta}}\langle H\rangle_{\boldsymbol{\theta}}$.
In our ansatz, the analytical expression of the gradient with respect to the parameter theta is provided by Eq.~\eqref{gradient}.
Therefore, we can calculate the energy gradient by tuning the circuit parameters to $\theta_i \pm \pi/2$, performing the experiments, and dealing with the data.

Our experiment procedure can be described by the following steps:  
\begin{itemize}
\item Randomly initialized the circuit parameter $\boldsymbol{\theta}$.
\item Change one of $\theta_i$ to $\theta_i+\pi/2$ and execute the circuit. Collect the bit strings for different Hamiltonian terms by measuring on a different basis. Perform readout error mitigation to the collect bit strings. Then we repeat for $\theta_i - \pi/2$. 
 \item Estimate the energy expectation value by assembling the statistics of all Hamiltonian terms.  Then calculate the gradient of all parameters via Eq.~\eqref{gradient}.  
\item Use gradient descent method to update the circuit parameters $\boldsymbol{\theta'}=\boldsymbol{\theta}-\alpha*\nabla E(\boldsymbol{\theta})$.  
\end{itemize}


This completes one training epoch. We measure and record the energy every interaction and stop training when a prescribed convergence criterion is met.

We carried out various two-qubit experiments on real quantum computers with different noise levels to compare our simulation results. The models include the transverse field Ising model, the transverse field Heisenberg model, and the Heisenberg model. Because the two-qubit gate error rate is crucial to our experimental results, we choose the CNOT error rate as the critical parameter to evaluate the quantum computers' noise level. The computers, therefore, are divided into three types:
\begin{itemize}
\item low noise type with CNOT error rate around 0.7\%, include $Valencia\{Q_0,Q_1\}$ and $Ourense\{Q_0,Q_1\}$.
\item intermediate noise type with  CNOT error rate around 1.4\%, include $Essex\{Q_1,Q_2\}$ and $Valencia\{Q_1,Q_3\}$.
\item high noise type with CNOT error rate around 3\%, include the $Yorktown\{Q_1,Q_2\}$.
\end{itemize}

Besides the gate noise, the readout errors are also crucial for IBMQ devices.
Here we consider the simplest linear algebra measurement error mitigation scheme. On every IBMQ device we do projective measurement  and obtain one of the strings $\{0,1\}^{\otimes 2}$. Through tomography of measurement process, we get the probability of string $S_j$ becoming $S_k$, denoted by $P_{kj}$. Suppose we repeat the same measurement many times and have the string probability distribution $C_{\text {noisy }}$, then
\begin{equation}
C_{\text {mitigated}}=P^{-1} C_{\text {noisy }}
\end{equation}
provides the probability distribution with measurement error mitigated, although $P^{-1}$ is not a physical operation. After measurement error mitigation, the result are mainly effected by the gate noise. This experiment process matches our simulation setup.

\begin{figure}[htb]
\centerline{\includegraphics[width=0.45\textwidth]{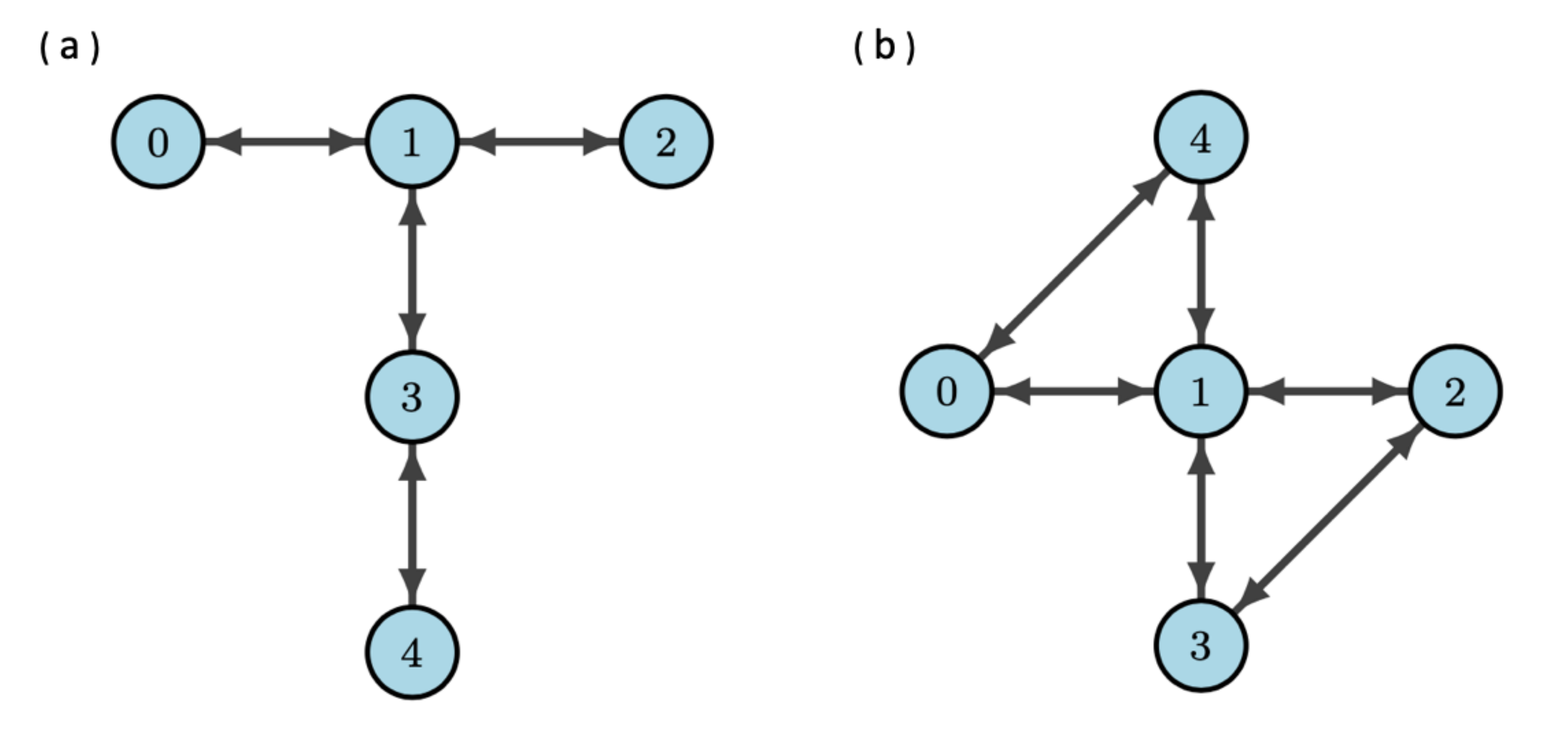}}
\caption{The physical structure of (a) Valencia, Ourense and Essex (b) Yorktown devices}
\label{fig:IBMStruct}
\end{figure}

\begin{table}[h!]
\centering
\addtolength{\tabcolsep}{2pt}
\caption{The experiment calibration parameters are supplied by the IBM Quantum Experience. The simulations use the corresponding average value. $T_1$ denotes amplitude damping time constant, $T_2$ denotes dephasing time constant, $\epsilon_{1q}$ denotes single-qubit gate error rate,  $\epsilon_{2q}$ denotes two-qubit gate error rate,  $ t_{1q} $ denotes single-qubit gate duration,  $ t_{2q} $ denotes two-qubit gate duration.}
\label{tab:IBMDepram}
{
\begin{tabular}{@{}lllllll@{}}
\toprule
  &$T_1 (\mu s)$     & $T_2 (\mu s)$ &  $\epsilon_{1q}$     & $\epsilon_{2q}$  &$ t_{1q} (ns)$ & $t_{2q}(ns)$ \\
\midrule
\textbf{Valencia} \\
Q0& 103.2 & 73.4 & 3.65E-4& 7.08E-3 &  35.6 & 263.1\\
Q1& 106.6 & 49.1 & 4.97E-4& 7.08E-3  & 35.6  & 298.7 \\
average &104.9 & 61.3  & 4.31E-4 & 7.08E-3 & 35.6 & 280.9 \\
\midrule
\textbf{Ourense}       \\
Q0& 122.5& 63.2& 2.80E-4& 6.07E-3 &35.6  &270.2\\
Q1& 96.6& 32.3& 3.57E-4& 6.07E-3   &35.6  &234.7 \\
average &109.6 & 47.8 & 3.19E-4 & 6.07E-3 & 35.6 & 252.5 \\
 \midrule
\textbf{Essex}     \\
Q1& 117.2& 139.2& 4.31E-4& 1.44E-2 &64.0 & 682.7\\
Q2& 100.1& 184.8& 5.41E-4& 1.44E-2&64.0 & 618.7 \\
average & 108.7 & 162.0 &4.86E-4 & 1.44E-2 & 64.0 & 650.7\\
 \midrule
\textbf{Valencia}      \\
Q1& 106.6& 49.1& 4.97E-4& 1.46E-2  & 35.6 & 540.4\\
Q3& 102.2& 47.1& 3.03E-4& 1.46E-2 & 35.6 & 576.0\\ 
average&104.4 & 48.1 &4.00E-4 & 1.46E-2 & 35.6 & 558.2\\
\midrule
\textbf{Yorktown}    \\
Q1& 61.2& 26.6& 1.27E-3& 3.48E-2 & 35.6  & 533.3 \\
Q2& 73.3& 76.8& 4.81E-4& 3.48E-2 & 35.6  & 568.9 \\
average& 67.3 & 51.7 & 3.04E-4& 3.48E-2 & 35.6 & 551.1\\
\bottomrule
\end{tabular}}
\end{table}

\subsection{Simulation method}
To compare with the realistic noise quantum device, we model the noise consisting of the single-qubit gate and two-qubit gate errors. The single-qubit gate error consists of a single qubit depolarizing error followed by a single qubit thermal relaxation error, and the two-qubit gate error consists of a two-qubit depolarizing error followed by a single-qubit thermal relaxation error on both qubits in the gate. The thermal relaxation error model applies the amplitude damping noise after dephasing noise in each one- or two-qubit gate. This  thermal relaxation error model is characterized through the paramters ($T_1, T_2,  t_{q}$) and the noise probability is formulated by
\begin{eqnarray}
p_{damping} &=& 1 - e^{-\frac{t_{q}}{T_1}}, \\
p_{dephasing} &=& \frac{1}{2}\left(1- e^{-2\gamma}\right), 
\end{eqnarray}
where $\gamma = \frac{t_q}{T_2} - \frac{t_q}{2T_1}$. When the thermal relaxation error model apply to single qubit gate, $t_{q} = {t_{1q}}$ and  $t_{q} = {t_{2q}}$ for two qubit gate. In Ref.~\cite{Ayral2020Quantum}, the probability of the depolarizing error $p_{1q}$ and $p_{2q}$ are then set so that the combined average process infidelity $\epsilon_{\text{avg}}^{1q}$ and $\epsilon_{\text{avg}}^{2q}$ from the depolarizing error followed by the thermal relaxation is equal to the gate calibration error rates  from IBM Quantum Experience. Here we just set $p_{1q} = \epsilon_{\text{avg}}^{1q}$ and $p_{2q}=\epsilon_{\text{avg}}^{2q}$ for simplicity.


Combining with the thermal relaxation error model and depolarizing noise model, the final noise model to approximate the noise of IBM quantum device is characterized by the parameters $\{T_1, T_2,  t_{1q}, t_{2q},  p_{1q},  p_{2q}\}$. Where $p_{1q}$ is the probability of one qubit depolarizing and $p_{2q}$  is the probability of a two-qubit depolarizing model. The value of the noise model to simulate different IBM Quantum Experience backend is given in Table~\ref{tab:IBMDepram}.

\subsection{Results}
The results are summarized in Table~\ref{tab:IBMExpValue}. The relative deviation between experiment and simulation is defined as $\left|E_{\text{exp}} - E_{\text{sim}} \right|/E_{\text{ext}}$, where $E_{\text{exp}}$ is the ground state energy obtained by IBM experiment, $E_{sim}$ is from our noisy VQE simulation and $E_{ext}$ is the exact ground state energy. 

\begin{table}[h!]
\centering
\addtolength{\tabcolsep}{2pt}
\caption{The comparison between the simulation of noise VQE and the IBM experiment. The relative deviation between experiment and simulation is defined as $\left|E_{\text{exp}} - E_{\text{sim}} \right|/E_{\text{ext}}$, where $E_{\text{exp}}$ is the ground state energy obtained by IBM experiment, $E_{\text{sim}}$ is from our noisy VQE simulation and $E_{\text{ext}}$ is the exact ground state energy.}
\label{tab:IBMExpValue}
{
\begin{tabular}{@{}lllllll@{}}
\toprule
                                       &$E_{\text{exp}}$  & $E_{\text{sim}}$ &  $E_{\text{ext}}$     & $\left|E_{\text{exp}} - E_{\text{sim}} \right|/E_{\text{ext}}$ \\
\midrule
\textbf{TIsing} \\
Valencia[0,1]   &  -2.31158  & -2.18543   &-2.23607 &  5.64\%\\
Valencia[1,3]   & -2.26336  & -2.13645   &-2.23607 & 5.68\% \\
Yorktown[1,2]   &-2.21259    & -2.07986   &-2.23607 & 5.93\%\\
 \midrule
\textbf{Heisenberg}\\
Ourense[0,1] &          -2.98321      & -2.87237 & -3.0 &  3.69\%\\
Essex[1,2] &            -2.99860        & -2.86530 & -3.0&  4.44\%\\
Yorktown[1,2]&       -2.93727        & -2.65497 & -3.0 &  9.41\%\\
\midrule
\textbf{THeisenberg} \\
 Valencia[0,1] &   -2.99999            & -2.87627 & -3.0 & 4.12\%\\
 Essex[1,2]     &   -2.94082             & -2.85353 & -3.0 & 2.91\%\\
Yorktown[1,2] &      -2.94834        & -2.63922  & -3.0 & 10.3\%\\
\bottomrule
\end{tabular}}
\end{table}

We find our noisy VQE simulation result agrees with the IBMQ experiment within 10\%. The deviation from the exact ground state increase as the noise level increase. We further analyze the trajectory of the cost function. The low noise backends tend to steadily converge to the minimum, while the high noise backends tend to oscillate around the minimal. That is because when noise possibility increases, the gradient we calculated via Eq.~\eqref{gradient} tends to randomly fluctuate around zero when the system is close to the ground state. It is worth mentioning that some of the $E_{\text{exp}}$s are lower than ground state energy. We perform measurements on changing density matrices when we collect bit strings for different hamiltonian terms due to the circuit noise. Therefore, it is possible to get energy lower than exact ground state energy when the ground state of hamiltonian $H$ is not the ground state of all the measured operators. We further carried out the $4$-qubit  VQE experiment with $2$ logical circuit depth on IBMQ. The results agree with the previous analysis. The minimal energy shows excellent consistency with simulation one. Similarly, the gradient oscillates randomly rather than gradually reduces to zero when searching around the minimum due to the noise effect.

\section{Discussion}
In this paper, we systematically studied the effect of noise on the VQE algorithm.  We simulate the noisy VQE with local noise models, including amplitude damping, dephasing, and depolarizing noise. We demonstrate that the VQE ground state energy gradually deviates from exact ground state energy as noise probability increases. By comparing different noise models, we find that depolarizing noise accumulates as circuit depth increases as one will typically expect. However, the dephasing and damping noise shows less noise accumulation for the models we study (see an explanation in Appendix A). To compare with the real quantum computer, we build a noise model for simulating the VQE experiments on IBMQ. The simulation results are consistent with the IBMQ experimental data.

We have shown that the VQE may suppress the accumulation of damping and dephasing noise in specific models. In principle, some other variational hybrid quantum-classical algorithms may also exhibit similar effects. It is also worth exploring the noise effect for other variational quantum algorithms. We also explore the quantumness of noisy VQE by comparing the noisy VQE simulation with the ground state problem's mean-field solution. We find that the VQE can outperform the mean-field solution of many-body ground state problems at the noise level of near-term quantum computers. These results suggest that the noisy quantum system can remain entanglement at the noise level of NISQ devices.

\section*{Acknowledgement}
We thank Dr. Y.-X. Liu for helpful discussions. PX thanks the support by the Key-Area Research and Development Program of Guangdong Province (No.2019B121204008). We acknowledge use of the IBM Q for this work. The views expressed are those of the authors and do not reflect the official policy or position of IBM or the IBM Q team.

\appendix
\section{The noise accumulation}

\begin{figure}[tbh]
\centerline{\includegraphics[width=0.50\textwidth]{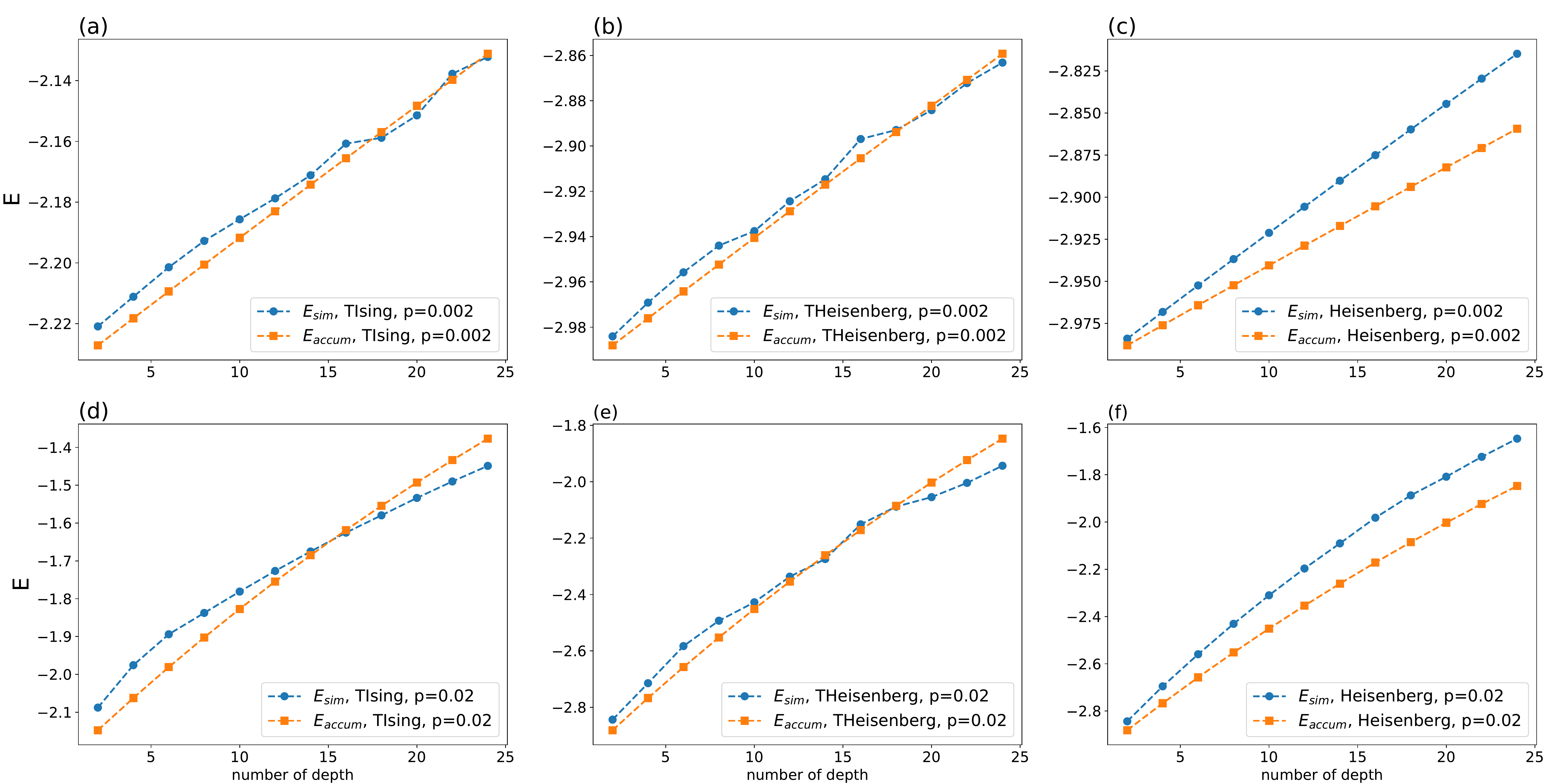}}
\caption{The accumulation phenomena of the local depolarizing noise in VQE. Given the local depolarizing parameter $p = 0.002$, the global depolarizing parameter $\epsilon$(y-axis) will increase in the quantum circuit depth (x-axis) increase in (a) the transverse field Ising model, (b) the transverse field Heisenberg model, and (c) Heisenberg model. The numerical result is also consistent with the analytical result. When the noise is large, say $p=0.02$, the conclusion holds in all three models (d-f).}
\label{fig:depolarizing_accumulation}
\end{figure}

The depolarizing noise accumulates as the circuit depth can be understood as follows.  When each gate just has depolarizing noise, we assume that the circuit output density has the same form
\begin{eqnarray}\label{eq:global_depolarizing}
\rho'  =   (1-\epsilon) U \rho U^{\dagger} +  \epsilon \frac{I}{2^n},
\end{eqnarray}
where $n$ is the number of qubit. $\epsilon$ is the total noise probability of the whole quantum circuit. The noisy VQE energy can be given by  
\begin{eqnarray}\label{eq:energy_relation}
E &=& \text{Tr}(H\rho') \nonumber \\ 
   &=& \text{Tr}\left(H(1-\epsilon)\rho_{0} +  H\epsilon\frac{I}{2^n}\right) \nonumber \\
   &=& (1-\epsilon) \text{Tr}(H\rho_{0}) + \frac{\epsilon}{2^n}\text{Tr}(H) \nonumber \\
   &=& (1-\epsilon) E_0 + \frac{\text{Tr}(H)}{2^n}\epsilon.
\end{eqnarray}
Assume  $\text{Tr}(H) = 0$, we have $E=(1-\epsilon) E_0 \geq E_{0}$, where $E_{0}$ is the exact ground state energy.

The relation between the local gate depolarizing parameter $p$ and the circuit depolarizing parameter $\epsilon$ can be approximated as 
\begin{eqnarray}
\rho'  \approx   (1-p)^d U \rho U^{\dagger} + (1 -(1-p)^d) \frac{I}{2^n},
\end{eqnarray} 
where $d$ is the depth of the quantum circuit. Given the local depolarizing parameter $p$, we can analytically derive the global depolarizing parameter
\begin{eqnarray}\label{eq:depolarizing_accumulate}
\epsilon \approx 1 - (1-p)^d.
\end{eqnarray} 
The simple approximation can explain the accumulation of depolarizing noise. Combine Eq.~\ref{eq:energy_relation} and Eq.~\ref{eq:depolarizing_accumulate}, we can calculate the analytical noise VQE energy $E_{\text{accum}}$, which is shown in the orange dot point line in Fig.~\ref{fig:depolarizing_accumulation}. We find that $E_{\text{accum}}$ is consistent with the simulated energy of noisy VQE $E_{\text{sim}}$, as shown in Fig.~\ref{fig:depolarizing_accumulation}.

The damping and dephasing noise in VQE do not accumulate as quantum circuit depth can be understood as follows.  The dephasing and amplitude can  be accumulated as circuit depth increases when applying in general density matrix expected the $| 0 \rangle ^{\otimes n}$. The VQE can adapt the parameters to make foregoing redundancy circuit layer to output $| 0 \rangle ^{\otimes n}$, which can not be affected by damping and dephasing noise. The parameters of the remaining circuit layers are optimized to get the minimal energy.

\bibliographystyle{apsrev4-1}
\bibliography{noiseVQE}

\begin{thebibliography}{57}%
\makeatletter
\providecommand \@ifxundefined [1]{%
 \@ifx{#1\undefined}
}%
\providecommand \@ifnum [1]{%
 \ifnum #1\expandafter \@firstoftwo
 \else \expandafter \@secondoftwo
 \fi
}%
\providecommand \@ifx [1]{%
 \ifx #1\expandafter \@firstoftwo
 \else \expandafter \@secondoftwo
 \fi
}%
\providecommand \natexlab [1]{#1}%
\providecommand \enquote  [1]{``#1''}%
\providecommand \bibnamefont  [1]{#1}%
\providecommand \bibfnamefont [1]{#1}%
\providecommand \citenamefont [1]{#1}%
\providecommand \href@noop [0]{\@secondoftwo}%
\providecommand \href [0]{\begingroup \@sanitize@url \@href}%
\providecommand \@href[1]{\@@startlink{#1}\@@href}%
\providecommand \@@href[1]{\endgroup#1\@@endlink}%
\providecommand \@sanitize@url [0]{\catcode `\\12\catcode `\$12\catcode
  `\&12\catcode `\#12\catcode `\^12\catcode `\_12\catcode `\%12\relax}%
\providecommand \@@startlink[1]{}%
\providecommand \@@endlink[0]{}%
\providecommand \url  [0]{\begingroup\@sanitize@url \@url }%
\providecommand \@url [1]{\endgroup\@href {#1}{\urlprefix }}%
\providecommand \urlprefix  [0]{URL }%
\providecommand \Eprint [0]{\href }%
\providecommand \doibase [0]{http://dx.doi.org/}%
\providecommand \selectlanguage [0]{\@gobble}%
\providecommand \bibinfo  [0]{\@secondoftwo}%
\providecommand \bibfield  [0]{\@secondoftwo}%
\providecommand \translation [1]{[#1]}%
\providecommand \BibitemOpen [0]{}%
\providecommand \bibitemStop [0]{}%
\providecommand \bibitemNoStop [0]{.\EOS\space}%
\providecommand \EOS [0]{\spacefactor3000\relax}%
\providecommand \BibitemShut  [1]{\csname bibitem#1\endcsname}%
\let\auto@bib@innerbib\@empty
\bibitem [{\citenamefont {Shor}(1999)}]{shor1999polynomial}%
  \BibitemOpen
  \bibfield  {author} {\bibinfo {author} {\bibfnamefont {P.~W.}\ \bibnamefont
  {Shor}},\ }\href {\doibase 10.1137/S0036144598347011} {\bibfield  {journal}
  {\bibinfo  {journal} {SIAM Review}\ }\textbf {\bibinfo {volume} {41}},\
  \bibinfo {pages} {303} (\bibinfo {year} {1999})}\BibitemShut {NoStop}%
\bibitem [{\citenamefont {Harrow}\ \emph {et~al.}(2009)\citenamefont {Harrow},
  \citenamefont {Hassidim},\ and\ \citenamefont {Lloyd}}]{harrow2009quantum}%
  \BibitemOpen
  \bibfield  {author} {\bibinfo {author} {\bibfnamefont {A.~W.}\ \bibnamefont
  {Harrow}}, \bibinfo {author} {\bibfnamefont {A.}~\bibnamefont {Hassidim}}, \
  and\ \bibinfo {author} {\bibfnamefont {S.}~\bibnamefont {Lloyd}},\ }\href
  {\doibase 10.1103/PhysRevLett.103.150502} {\bibfield  {journal} {\bibinfo
  {journal} {Phys. Rev. Lett.}\ }\textbf {\bibinfo {volume} {103}},\ \bibinfo
  {pages} {150502} (\bibinfo {year} {2009})}\BibitemShut {NoStop}%
\bibitem [{\citenamefont {Shor}(1995)}]{Shor1995Scheme}%
  \BibitemOpen
  \bibfield  {author} {\bibinfo {author} {\bibfnamefont {P.~W.}\ \bibnamefont
  {Shor}},\ }\href {\doibase 10.1103/PhysRevA.52.R2493} {\bibfield  {journal}
  {\bibinfo  {journal} {Phys. Rev. A}\ }\textbf {\bibinfo {volume} {52}},\
  \bibinfo {pages} {R2493} (\bibinfo {year} {1995})}\BibitemShut {NoStop}%
\bibitem [{\citenamefont {Steane}(1996)}]{Steane1996Error}%
  \BibitemOpen
  \bibfield  {author} {\bibinfo {author} {\bibfnamefont {A.~M.}\ \bibnamefont
  {Steane}},\ }\href {\doibase 10.1103/PhysRevLett.77.793} {\bibfield
  {journal} {\bibinfo  {journal} {Phys. Rev. Lett.}\ }\textbf {\bibinfo
  {volume} {77}},\ \bibinfo {pages} {793} (\bibinfo {year} {1996})}\BibitemShut
  {NoStop}%
\bibitem [{\citenamefont {Gottesman}(1996)}]{Gottesman1996Class}%
  \BibitemOpen
  \bibfield  {author} {\bibinfo {author} {\bibfnamefont {D.}~\bibnamefont
  {Gottesman}},\ }\href {\doibase 10.1103/PhysRevA.54.1862} {\bibfield
  {journal} {\bibinfo  {journal} {Phys. Rev. A}\ }\textbf {\bibinfo {volume}
  {54}},\ \bibinfo {pages} {1862} (\bibinfo {year} {1996})}\BibitemShut
  {NoStop}%
\bibitem [{\citenamefont {Bennett}\ \emph {et~al.}(1996)\citenamefont
  {Bennett}, \citenamefont {DiVincenzo}, \citenamefont {Smolin},\ and\
  \citenamefont {Wootters}}]{Bennett1996Mixed}%
  \BibitemOpen
  \bibfield  {author} {\bibinfo {author} {\bibfnamefont {C.~H.}\ \bibnamefont
  {Bennett}}, \bibinfo {author} {\bibfnamefont {D.~P.}\ \bibnamefont
  {DiVincenzo}}, \bibinfo {author} {\bibfnamefont {J.~A.}\ \bibnamefont
  {Smolin}}, \ and\ \bibinfo {author} {\bibfnamefont {W.~K.}\ \bibnamefont
  {Wootters}},\ }\href {\doibase 10.1103/PhysRevA.54.3824} {\bibfield
  {journal} {\bibinfo  {journal} {Phys. Rev. A}\ }\textbf {\bibinfo {volume}
  {54}},\ \bibinfo {pages} {3824} (\bibinfo {year} {1996})}\BibitemShut
  {NoStop}%
\bibitem [{\citenamefont {Knill}\ and\ \citenamefont
  {Laflamme}(1997)}]{Knill1997Theory}%
  \BibitemOpen
  \bibfield  {author} {\bibinfo {author} {\bibfnamefont {E.}~\bibnamefont
  {Knill}}\ and\ \bibinfo {author} {\bibfnamefont {R.}~\bibnamefont
  {Laflamme}},\ }\href {\doibase 10.1103/PhysRevA.55.900} {\bibfield  {journal}
  {\bibinfo  {journal} {Phys. Rev. A}\ }\textbf {\bibinfo {volume} {55}},\
  \bibinfo {pages} {900} (\bibinfo {year} {1997})}\BibitemShut {NoStop}%
\bibitem [{\citenamefont {Gaitan}(2008)}]{gaitan2008quantum}%
  \BibitemOpen
  \bibfield  {author} {\bibinfo {author} {\bibfnamefont {F.}~\bibnamefont
  {Gaitan}},\ }\href@noop {} {\emph {\bibinfo {title} {Quantum error correction
  and fault tolerant quantum computing}}}\ (\bibinfo  {publisher} {CRC Press},\
  \bibinfo {year} {2008})\BibitemShut {NoStop}%
\bibitem [{\citenamefont {Preskill}(2018)}]{preskill2018quantum}%
  \BibitemOpen
  \bibfield  {author} {\bibinfo {author} {\bibfnamefont {J.}~\bibnamefont
  {Preskill}},\ }\href {\doibase 10.22331/q-2018-08-06-79} {\bibfield
  {journal} {\bibinfo  {journal} {{Quantum}}\ }\textbf {\bibinfo {volume}
  {2}},\ \bibinfo {pages} {79} (\bibinfo {year} {2018})}\BibitemShut {NoStop}%
\bibitem [{\citenamefont {Arute}\ \emph {et~al.}(2019)\citenamefont {Arute},
  \citenamefont {Arya}, \citenamefont {Babbush}, \citenamefont {Bacon},
  \citenamefont {Bardin}, \citenamefont {Barends}, \citenamefont {Biswas},
  \citenamefont {Boixo}, \citenamefont {Brandao}, \citenamefont {Buell},
  \citenamefont {Burkett}, \citenamefont {Chen}, \citenamefont {Chen},
  \citenamefont {Chiaro}, \citenamefont {Collins}, \citenamefont {Courtney},
  \citenamefont {Dunsworth}, \citenamefont {Farhi}, \citenamefont {Foxen},
  \citenamefont {Fowler}, \citenamefont {Gidney}, \citenamefont {Giustina},
  \citenamefont {Graff}, \citenamefont {Guerin}, \citenamefont {Habegger},
  \citenamefont {Harrigan}, \citenamefont {Hartmann}, \citenamefont {Ho},
  \citenamefont {Hoffmann}, \citenamefont {Huang}, \citenamefont {Humble},
  \citenamefont {Isakov}, \citenamefont {Jeffrey}, \citenamefont {Jiang},
  \citenamefont {Kafri}, \citenamefont {Kechedzhi}, \citenamefont {Kelly},
  \citenamefont {Klimov}, \citenamefont {Knysh}, \citenamefont {Korotkov},
  \citenamefont {Kostritsa}, \citenamefont {Landhuis}, \citenamefont
  {Lindmark}, \citenamefont {Lucero}, \citenamefont {Lyakh}, \citenamefont
  {Mandr{\`a}}, \citenamefont {McClean}, \citenamefont {McEwen}, \citenamefont
  {Megrant}, \citenamefont {Mi}, \citenamefont {Michielsen}, \citenamefont
  {Mohseni}, \citenamefont {Mutus}, \citenamefont {Naaman}, \citenamefont
  {Neeley}, \citenamefont {Neill}, \citenamefont {Niu}, \citenamefont {Ostby},
  \citenamefont {Petukhov}, \citenamefont {Platt}, \citenamefont {Quintana},
  \citenamefont {Rieffel}, \citenamefont {Roushan}, \citenamefont {Rubin},
  \citenamefont {Sank}, \citenamefont {Satzinger}, \citenamefont {Smelyanskiy},
  \citenamefont {Sung}, \citenamefont {Trevithick}, \citenamefont
  {Vainsencher}, \citenamefont {Villalonga}, \citenamefont {White},
  \citenamefont {Yao}, \citenamefont {Yeh}, \citenamefont {Zalcman},
  \citenamefont {Neven},\ and\ \citenamefont {Martinis}}]{Google2019Supremacy}%
  \BibitemOpen
  \bibfield  {author} {\bibinfo {author} {\bibfnamefont {F.}~\bibnamefont
  {Arute}}, \bibinfo {author} {\bibfnamefont {K.}~\bibnamefont {Arya}},
  \bibinfo {author} {\bibfnamefont {R.}~\bibnamefont {Babbush}}, \bibinfo
  {author} {\bibfnamefont {D.}~\bibnamefont {Bacon}}, \bibinfo {author}
  {\bibfnamefont {J.~C.}\ \bibnamefont {Bardin}}, \bibinfo {author}
  {\bibfnamefont {R.}~\bibnamefont {Barends}}, \bibinfo {author} {\bibfnamefont
  {R.}~\bibnamefont {Biswas}}, \bibinfo {author} {\bibfnamefont
  {S.}~\bibnamefont {Boixo}}, \bibinfo {author} {\bibfnamefont {F.~G. S.~L.}\
  \bibnamefont {Brandao}}, \bibinfo {author} {\bibfnamefont {D.~A.}\
  \bibnamefont {Buell}}, \bibinfo {author} {\bibfnamefont {B.}~\bibnamefont
  {Burkett}}, \bibinfo {author} {\bibfnamefont {Y.}~\bibnamefont {Chen}},
  \bibinfo {author} {\bibfnamefont {Z.}~\bibnamefont {Chen}}, \bibinfo {author}
  {\bibfnamefont {B.}~\bibnamefont {Chiaro}}, \bibinfo {author} {\bibfnamefont
  {R.}~\bibnamefont {Collins}}, \bibinfo {author} {\bibfnamefont
  {W.}~\bibnamefont {Courtney}}, \bibinfo {author} {\bibfnamefont
  {A.}~\bibnamefont {Dunsworth}}, \bibinfo {author} {\bibfnamefont
  {E.}~\bibnamefont {Farhi}}, \bibinfo {author} {\bibfnamefont
  {B.}~\bibnamefont {Foxen}}, \bibinfo {author} {\bibfnamefont
  {A.}~\bibnamefont {Fowler}}, \bibinfo {author} {\bibfnamefont
  {C.}~\bibnamefont {Gidney}}, \bibinfo {author} {\bibfnamefont
  {M.}~\bibnamefont {Giustina}}, \bibinfo {author} {\bibfnamefont
  {R.}~\bibnamefont {Graff}}, \bibinfo {author} {\bibfnamefont
  {K.}~\bibnamefont {Guerin}}, \bibinfo {author} {\bibfnamefont
  {S.}~\bibnamefont {Habegger}}, \bibinfo {author} {\bibfnamefont {M.~P.}\
  \bibnamefont {Harrigan}}, \bibinfo {author} {\bibfnamefont {M.~J.}\
  \bibnamefont {Hartmann}}, \bibinfo {author} {\bibfnamefont {A.}~\bibnamefont
  {Ho}}, \bibinfo {author} {\bibfnamefont {M.}~\bibnamefont {Hoffmann}},
  \bibinfo {author} {\bibfnamefont {T.}~\bibnamefont {Huang}}, \bibinfo
  {author} {\bibfnamefont {T.~S.}\ \bibnamefont {Humble}}, \bibinfo {author}
  {\bibfnamefont {S.~V.}\ \bibnamefont {Isakov}}, \bibinfo {author}
  {\bibfnamefont {E.}~\bibnamefont {Jeffrey}}, \bibinfo {author} {\bibfnamefont
  {Z.}~\bibnamefont {Jiang}}, \bibinfo {author} {\bibfnamefont
  {D.}~\bibnamefont {Kafri}}, \bibinfo {author} {\bibfnamefont
  {K.}~\bibnamefont {Kechedzhi}}, \bibinfo {author} {\bibfnamefont
  {J.}~\bibnamefont {Kelly}}, \bibinfo {author} {\bibfnamefont {P.~V.}\
  \bibnamefont {Klimov}}, \bibinfo {author} {\bibfnamefont {S.}~\bibnamefont
  {Knysh}}, \bibinfo {author} {\bibfnamefont {A.}~\bibnamefont {Korotkov}},
  \bibinfo {author} {\bibfnamefont {F.}~\bibnamefont {Kostritsa}}, \bibinfo
  {author} {\bibfnamefont {D.}~\bibnamefont {Landhuis}}, \bibinfo {author}
  {\bibfnamefont {M.}~\bibnamefont {Lindmark}}, \bibinfo {author}
  {\bibfnamefont {E.}~\bibnamefont {Lucero}}, \bibinfo {author} {\bibfnamefont
  {D.}~\bibnamefont {Lyakh}}, \bibinfo {author} {\bibfnamefont
  {S.}~\bibnamefont {Mandr{\`a}}}, \bibinfo {author} {\bibfnamefont {J.~R.}\
  \bibnamefont {McClean}}, \bibinfo {author} {\bibfnamefont {M.}~\bibnamefont
  {McEwen}}, \bibinfo {author} {\bibfnamefont {A.}~\bibnamefont {Megrant}},
  \bibinfo {author} {\bibfnamefont {X.}~\bibnamefont {Mi}}, \bibinfo {author}
  {\bibfnamefont {K.}~\bibnamefont {Michielsen}}, \bibinfo {author}
  {\bibfnamefont {M.}~\bibnamefont {Mohseni}}, \bibinfo {author} {\bibfnamefont
  {J.}~\bibnamefont {Mutus}}, \bibinfo {author} {\bibfnamefont
  {O.}~\bibnamefont {Naaman}}, \bibinfo {author} {\bibfnamefont
  {M.}~\bibnamefont {Neeley}}, \bibinfo {author} {\bibfnamefont
  {C.}~\bibnamefont {Neill}}, \bibinfo {author} {\bibfnamefont {M.~Y.}\
  \bibnamefont {Niu}}, \bibinfo {author} {\bibfnamefont {E.}~\bibnamefont
  {Ostby}}, \bibinfo {author} {\bibfnamefont {A.}~\bibnamefont {Petukhov}},
  \bibinfo {author} {\bibfnamefont {J.~C.}\ \bibnamefont {Platt}}, \bibinfo
  {author} {\bibfnamefont {C.}~\bibnamefont {Quintana}}, \bibinfo {author}
  {\bibfnamefont {E.~G.}\ \bibnamefont {Rieffel}}, \bibinfo {author}
  {\bibfnamefont {P.}~\bibnamefont {Roushan}}, \bibinfo {author} {\bibfnamefont
  {N.~C.}\ \bibnamefont {Rubin}}, \bibinfo {author} {\bibfnamefont
  {D.}~\bibnamefont {Sank}}, \bibinfo {author} {\bibfnamefont {K.~J.}\
  \bibnamefont {Satzinger}}, \bibinfo {author} {\bibfnamefont {V.}~\bibnamefont
  {Smelyanskiy}}, \bibinfo {author} {\bibfnamefont {K.~J.}\ \bibnamefont
  {Sung}}, \bibinfo {author} {\bibfnamefont {M.~D.}\ \bibnamefont
  {Trevithick}}, \bibinfo {author} {\bibfnamefont {A.}~\bibnamefont
  {Vainsencher}}, \bibinfo {author} {\bibfnamefont {B.}~\bibnamefont
  {Villalonga}}, \bibinfo {author} {\bibfnamefont {T.}~\bibnamefont {White}},
  \bibinfo {author} {\bibfnamefont {Z.~J.}\ \bibnamefont {Yao}}, \bibinfo
  {author} {\bibfnamefont {P.}~\bibnamefont {Yeh}}, \bibinfo {author}
  {\bibfnamefont {A.}~\bibnamefont {Zalcman}}, \bibinfo {author} {\bibfnamefont
  {H.}~\bibnamefont {Neven}}, \ and\ \bibinfo {author} {\bibfnamefont {J.~M.}\
  \bibnamefont {Martinis}},\ }\href {\doibase 10.1038/s41586-019-1666-5}
  {\bibfield  {journal} {\bibinfo  {journal} {Nature}\ }\textbf {\bibinfo
  {volume} {574}},\ \bibinfo {pages} {505} (\bibinfo {year}
  {2019})}\BibitemShut {NoStop}%
\bibitem [{\citenamefont {Zhong}\ \emph {et~al.}(2020)\citenamefont {Zhong},
  \citenamefont {Wang}, \citenamefont {Deng}, \citenamefont {Chen},
  \citenamefont {Peng}, \citenamefont {Luo}, \citenamefont {Qin}, \citenamefont
  {Wu}, \citenamefont {Ding}, \citenamefont {Hu}, \citenamefont {Hu},
  \citenamefont {Yang}, \citenamefont {Zhang}, \citenamefont {Li},
  \citenamefont {Li}, \citenamefont {Jiang}, \citenamefont {Gan}, \citenamefont
  {Yang}, \citenamefont {You}, \citenamefont {Wang}, \citenamefont {Li},
  \citenamefont {Liu}, \citenamefont {Lu},\ and\ \citenamefont
  {Pan}}]{Zhong2020Advantage}%
  \BibitemOpen
  \bibfield  {author} {\bibinfo {author} {\bibfnamefont {H.-S.}\ \bibnamefont
  {Zhong}}, \bibinfo {author} {\bibfnamefont {H.}~\bibnamefont {Wang}},
  \bibinfo {author} {\bibfnamefont {Y.-H.}\ \bibnamefont {Deng}}, \bibinfo
  {author} {\bibfnamefont {M.-C.}\ \bibnamefont {Chen}}, \bibinfo {author}
  {\bibfnamefont {L.-C.}\ \bibnamefont {Peng}}, \bibinfo {author}
  {\bibfnamefont {Y.-H.}\ \bibnamefont {Luo}}, \bibinfo {author} {\bibfnamefont
  {J.}~\bibnamefont {Qin}}, \bibinfo {author} {\bibfnamefont {D.}~\bibnamefont
  {Wu}}, \bibinfo {author} {\bibfnamefont {X.}~\bibnamefont {Ding}}, \bibinfo
  {author} {\bibfnamefont {Y.}~\bibnamefont {Hu}}, \bibinfo {author}
  {\bibfnamefont {P.}~\bibnamefont {Hu}}, \bibinfo {author} {\bibfnamefont
  {X.-Y.}\ \bibnamefont {Yang}}, \bibinfo {author} {\bibfnamefont {W.-J.}\
  \bibnamefont {Zhang}}, \bibinfo {author} {\bibfnamefont {H.}~\bibnamefont
  {Li}}, \bibinfo {author} {\bibfnamefont {Y.}~\bibnamefont {Li}}, \bibinfo
  {author} {\bibfnamefont {X.}~\bibnamefont {Jiang}}, \bibinfo {author}
  {\bibfnamefont {L.}~\bibnamefont {Gan}}, \bibinfo {author} {\bibfnamefont
  {G.}~\bibnamefont {Yang}}, \bibinfo {author} {\bibfnamefont {L.}~\bibnamefont
  {You}}, \bibinfo {author} {\bibfnamefont {Z.}~\bibnamefont {Wang}}, \bibinfo
  {author} {\bibfnamefont {L.}~\bibnamefont {Li}}, \bibinfo {author}
  {\bibfnamefont {N.-L.}\ \bibnamefont {Liu}}, \bibinfo {author} {\bibfnamefont
  {C.-Y.}\ \bibnamefont {Lu}}, \ and\ \bibinfo {author} {\bibfnamefont {J.-W.}\
  \bibnamefont {Pan}},\ }\href {\doibase 10.1126/science.abe8770} {\bibfield
  {journal} {\bibinfo  {journal} {Science}\ } (\bibinfo {year} {2020}),\
  10.1126/science.abe8770}\BibitemShut {NoStop}%
\bibitem [{\citenamefont {Peruzzo}\ \emph {et~al.}(2014)\citenamefont
  {Peruzzo}, \citenamefont {McClean}, \citenamefont {Shadbolt}, \citenamefont
  {Yung}, \citenamefont {Zhou}, \citenamefont {Love}, \citenamefont
  {Aspuru-Guzik},\ and\ \citenamefont {O’brien}}]{Peruzzo2014Variational}%
  \BibitemOpen
  \bibfield  {author} {\bibinfo {author} {\bibfnamefont {A.}~\bibnamefont
  {Peruzzo}}, \bibinfo {author} {\bibfnamefont {J.}~\bibnamefont {McClean}},
  \bibinfo {author} {\bibfnamefont {P.}~\bibnamefont {Shadbolt}}, \bibinfo
  {author} {\bibfnamefont {M.-H.}\ \bibnamefont {Yung}}, \bibinfo {author}
  {\bibfnamefont {X.-Q.}\ \bibnamefont {Zhou}}, \bibinfo {author}
  {\bibfnamefont {P.~J.}\ \bibnamefont {Love}}, \bibinfo {author}
  {\bibfnamefont {A.}~\bibnamefont {Aspuru-Guzik}}, \ and\ \bibinfo {author}
  {\bibfnamefont {J.~L.}\ \bibnamefont {O’brien}},\ }\href
  {https://www.nature.com/articles/ncomms5213} {\bibfield  {journal} {\bibinfo
  {journal} {Nature communications}\ }\textbf {\bibinfo {volume} {5}},\
  \bibinfo {pages} {4213} (\bibinfo {year} {2014})}\BibitemShut {NoStop}%
\bibitem [{\citenamefont {McClean}\ \emph {et~al.}(2016)\citenamefont
  {McClean}, \citenamefont {Romero}, \citenamefont {Babbush},\ and\
  \citenamefont {Aspuru-Guzik}}]{McClean2016VQE}%
  \BibitemOpen
  \bibfield  {author} {\bibinfo {author} {\bibfnamefont {J.~R.}\ \bibnamefont
  {McClean}}, \bibinfo {author} {\bibfnamefont {J.}~\bibnamefont {Romero}},
  \bibinfo {author} {\bibfnamefont {R.}~\bibnamefont {Babbush}}, \ and\
  \bibinfo {author} {\bibfnamefont {A.}~\bibnamefont {Aspuru-Guzik}},\ }\href
  {http://stacks.iop.org/1367-2630/18/i=2/a=023023} {\bibfield  {journal}
  {\bibinfo  {journal} {New Journal of Physics}\ }\textbf {\bibinfo {volume}
  {18}},\ \bibinfo {pages} {023023} (\bibinfo {year} {2016})}\BibitemShut
  {NoStop}%
\bibitem [{\citenamefont {Kandala}\ \emph {et~al.}(2017)\citenamefont
  {Kandala}, \citenamefont {Mezzacapo}, \citenamefont {Temme}, \citenamefont
  {Takita}, \citenamefont {Brink}, \citenamefont {Chow},\ and\ \citenamefont
  {Gambetta}}]{Kandala2017}%
  \BibitemOpen
  \bibfield  {author} {\bibinfo {author} {\bibfnamefont {A.}~\bibnamefont
  {Kandala}}, \bibinfo {author} {\bibfnamefont {A.}~\bibnamefont {Mezzacapo}},
  \bibinfo {author} {\bibfnamefont {K.}~\bibnamefont {Temme}}, \bibinfo
  {author} {\bibfnamefont {M.}~\bibnamefont {Takita}}, \bibinfo {author}
  {\bibfnamefont {M.}~\bibnamefont {Brink}}, \bibinfo {author} {\bibfnamefont
  {J.~M.}\ \bibnamefont {Chow}}, \ and\ \bibinfo {author} {\bibfnamefont
  {J.~M.}\ \bibnamefont {Gambetta}},\ }\href
  {https://www.nature.com/articles/nature23879} {\bibfield  {journal} {\bibinfo
   {journal} {Nature}\ }\textbf {\bibinfo {volume} {549}},\ \bibinfo {pages}
  {242} (\bibinfo {year} {2017})}\BibitemShut {NoStop}%
\bibitem [{\citenamefont {Farhi}\ \emph {et~al.}(2014)\citenamefont {Farhi},
  \citenamefont {Goldstone},\ and\ \citenamefont {Gutmann}}]{farhi2014quantum}%
  \BibitemOpen
  \bibfield  {author} {\bibinfo {author} {\bibfnamefont {E.}~\bibnamefont
  {Farhi}}, \bibinfo {author} {\bibfnamefont {J.}~\bibnamefont {Goldstone}}, \
  and\ \bibinfo {author} {\bibfnamefont {S.}~\bibnamefont {Gutmann}},\ }\href
  {https://arxiv.org/abs/1411.4028} {\bibfield  {journal} {\bibinfo  {journal}
  {arXiv:1411.4028}\ } (\bibinfo {year} {2014})}\BibitemShut {NoStop}%
\bibitem [{\citenamefont {Motta}\ \emph {et~al.}(2020)\citenamefont {Motta},
  \citenamefont {Sun}, \citenamefont {Tan}, \citenamefont {O’Rourke},
  \citenamefont {Ye}, \citenamefont {Minnich}, \citenamefont {Brand{\~a}o},\
  and\ \citenamefont {Chan}}]{motta2020determining}%
  \BibitemOpen
  \bibfield  {author} {\bibinfo {author} {\bibfnamefont {M.}~\bibnamefont
  {Motta}}, \bibinfo {author} {\bibfnamefont {C.}~\bibnamefont {Sun}}, \bibinfo
  {author} {\bibfnamefont {A.~T.}\ \bibnamefont {Tan}}, \bibinfo {author}
  {\bibfnamefont {M.~J.}\ \bibnamefont {O’Rourke}}, \bibinfo {author}
  {\bibfnamefont {E.}~\bibnamefont {Ye}}, \bibinfo {author} {\bibfnamefont
  {A.~J.}\ \bibnamefont {Minnich}}, \bibinfo {author} {\bibfnamefont {F.~G.}\
  \bibnamefont {Brand{\~a}o}}, \ and\ \bibinfo {author} {\bibfnamefont
  {G.~K.-L.}\ \bibnamefont {Chan}},\ }\href
  {https://doi.org/10.1038/s41567-019-0704-4} {\bibfield  {journal} {\bibinfo
  {journal} {Nature Physics}\ }\textbf {\bibinfo {volume} {16}},\ \bibinfo
  {pages} {205} (\bibinfo {year} {2020})}\BibitemShut {NoStop}%
\bibitem [{\citenamefont {McArdle}\ \emph {et~al.}(2019)\citenamefont
  {McArdle}, \citenamefont {Jones}, \citenamefont {Endo}, \citenamefont {Li},
  \citenamefont {Benjamin},\ and\ \citenamefont {Yuan}}]{McArdle2019QITE}%
  \BibitemOpen
  \bibfield  {author} {\bibinfo {author} {\bibfnamefont {S.}~\bibnamefont
  {McArdle}}, \bibinfo {author} {\bibfnamefont {T.}~\bibnamefont {Jones}},
  \bibinfo {author} {\bibfnamefont {S.}~\bibnamefont {Endo}}, \bibinfo {author}
  {\bibfnamefont {Y.}~\bibnamefont {Li}}, \bibinfo {author} {\bibfnamefont
  {S.~C.}\ \bibnamefont {Benjamin}}, \ and\ \bibinfo {author} {\bibfnamefont
  {X.}~\bibnamefont {Yuan}},\ }\href {\doibase 10.1038/s41534-019-0187-2}
  {\bibfield  {journal} {\bibinfo  {journal} {npj Quantum Information}\
  }\textbf {\bibinfo {volume} {5}},\ \bibinfo {pages} {75} (\bibinfo {year}
  {2019})}\BibitemShut {NoStop}%
\bibitem [{\citenamefont {Zoufal}\ \emph {et~al.}(2020)\citenamefont {Zoufal},
  \citenamefont {Lucchi},\ and\ \citenamefont
  {Woerner}}]{zoufal2020variational}%
  \BibitemOpen
  \bibfield  {author} {\bibinfo {author} {\bibfnamefont {C.}~\bibnamefont
  {Zoufal}}, \bibinfo {author} {\bibfnamefont {A.}~\bibnamefont {Lucchi}}, \
  and\ \bibinfo {author} {\bibfnamefont {S.}~\bibnamefont {Woerner}},\ }\href
  {https://arxiv.org/abs/2006.06004} {\bibfield  {journal} {\bibinfo  {journal}
  {arXiv:2006.06004}\ } (\bibinfo {year} {2020})}\BibitemShut {NoStop}%
\bibitem [{\citenamefont {Shingu}\ \emph {et~al.}(2020)\citenamefont {Shingu},
  \citenamefont {Seki}, \citenamefont {Watabe}, \citenamefont {Endo},
  \citenamefont {Matsuzaki}, \citenamefont {Kawabata}, \citenamefont {Nikuni},\
  and\ \citenamefont {Hakoshima}}]{shingu2020boltzmann}%
  \BibitemOpen
  \bibfield  {author} {\bibinfo {author} {\bibfnamefont {Y.}~\bibnamefont
  {Shingu}}, \bibinfo {author} {\bibfnamefont {Y.}~\bibnamefont {Seki}},
  \bibinfo {author} {\bibfnamefont {S.}~\bibnamefont {Watabe}}, \bibinfo
  {author} {\bibfnamefont {S.}~\bibnamefont {Endo}}, \bibinfo {author}
  {\bibfnamefont {Y.}~\bibnamefont {Matsuzaki}}, \bibinfo {author}
  {\bibfnamefont {S.}~\bibnamefont {Kawabata}}, \bibinfo {author}
  {\bibfnamefont {T.}~\bibnamefont {Nikuni}}, \ and\ \bibinfo {author}
  {\bibfnamefont {H.}~\bibnamefont {Hakoshima}},\ }\href
  {https://arxiv.org/abs/2007.00876} {\bibfield  {journal} {\bibinfo  {journal}
  {arXiv:2007.00876}\ } (\bibinfo {year} {2020})}\BibitemShut {NoStop}%
\bibitem [{\citenamefont {Wang}\ \emph {et~al.}(2018)\citenamefont {Wang},
  \citenamefont {Li}, \citenamefont {Yin},\ and\ \citenamefont
  {Zeng}}]{Wang2018IBM}%
  \BibitemOpen
  \bibfield  {author} {\bibinfo {author} {\bibfnamefont {Y.}~\bibnamefont
  {Wang}}, \bibinfo {author} {\bibfnamefont {Y.}~\bibnamefont {Li}}, \bibinfo
  {author} {\bibfnamefont {Z.-q.}\ \bibnamefont {Yin}}, \ and\ \bibinfo
  {author} {\bibfnamefont {B.}~\bibnamefont {Zeng}},\ }\href {\doibase
  10.1038/s41534-018-0095-x} {\bibfield  {journal} {\bibinfo  {journal} {npj
  Quantum Information}\ }\textbf {\bibinfo {volume} {4}},\ \bibinfo {pages}
  {46} (\bibinfo {year} {2018})}\BibitemShut {NoStop}%
\bibitem [{\citenamefont {Mooney}\ \emph {et~al.}(2021)\citenamefont {Mooney},
  \citenamefont {White}, \citenamefont {Hill},\ and\ \citenamefont
  {Hollenberg}}]{mooney2021wholedevice}%
  \BibitemOpen
  \bibfield  {author} {\bibinfo {author} {\bibfnamefont {G.~J.}\ \bibnamefont
  {Mooney}}, \bibinfo {author} {\bibfnamefont {G.~A.~L.}\ \bibnamefont
  {White}}, \bibinfo {author} {\bibfnamefont {C.~D.}\ \bibnamefont {Hill}}, \
  and\ \bibinfo {author} {\bibfnamefont {L.~C.~L.}\ \bibnamefont
  {Hollenberg}},\ }\href {https://arxiv.org/abs/2102.11521} {\bibfield
  {journal} {\bibinfo  {journal} {arXiv:2102.11521}\ } (\bibinfo {year}
  {2021})}\BibitemShut {NoStop}%
\bibitem [{\citenamefont {Xue}\ \emph {et~al.}(2019)\citenamefont {Xue},
  \citenamefont {Chen}, \citenamefont {Wu},\ and\ \citenamefont
  {Guo}}]{xue2019effects}%
  \BibitemOpen
  \bibfield  {author} {\bibinfo {author} {\bibfnamefont {C.}~\bibnamefont
  {Xue}}, \bibinfo {author} {\bibfnamefont {Z.-Y.}\ \bibnamefont {Chen}},
  \bibinfo {author} {\bibfnamefont {Y.-C.}\ \bibnamefont {Wu}}, \ and\ \bibinfo
  {author} {\bibfnamefont {G.-P.}\ \bibnamefont {Guo}},\ }\href
  {http://arxiv.org/abs/1909.02196} {\bibfield  {journal} {\bibinfo  {journal}
  {arXiv:1909.02196}\ } (\bibinfo {year} {2019})}\BibitemShut {NoStop}%
\bibitem [{\citenamefont {Marshall}\ \emph {et~al.}(2020)\citenamefont
  {Marshall}, \citenamefont {Wudarski}, \citenamefont {Hadfield},\ and\
  \citenamefont {Hogg}}]{Marshall2020Characterizing}%
  \BibitemOpen
  \bibfield  {author} {\bibinfo {author} {\bibfnamefont {J.}~\bibnamefont
  {Marshall}}, \bibinfo {author} {\bibfnamefont {F.}~\bibnamefont {Wudarski}},
  \bibinfo {author} {\bibfnamefont {S.}~\bibnamefont {Hadfield}}, \ and\
  \bibinfo {author} {\bibfnamefont {T.}~\bibnamefont {Hogg}},\ }\href {\doibase
  10.1088/2633-1357/abb0d7} {\bibfield  {journal} {\bibinfo  {journal} {{IOP}
  {SciNotes}}\ }\textbf {\bibinfo {volume} {1}},\ \bibinfo {pages} {025208}
  (\bibinfo {year} {2020})}\BibitemShut {NoStop}%
\bibitem [{\citenamefont {Wang}\ \emph
  {et~al.}(2020{\natexlab{a}})\citenamefont {Wang}, \citenamefont {Fontana},
  \citenamefont {Cerezo}, \citenamefont {Sharma}, \citenamefont {Sone},
  \citenamefont {Cincio},\ and\ \citenamefont {Coles}}]{wang2020noise}%
  \BibitemOpen
  \bibfield  {author} {\bibinfo {author} {\bibfnamefont {S.}~\bibnamefont
  {Wang}}, \bibinfo {author} {\bibfnamefont {E.}~\bibnamefont {Fontana}},
  \bibinfo {author} {\bibfnamefont {M.}~\bibnamefont {Cerezo}}, \bibinfo
  {author} {\bibfnamefont {K.}~\bibnamefont {Sharma}}, \bibinfo {author}
  {\bibfnamefont {A.}~\bibnamefont {Sone}}, \bibinfo {author} {\bibfnamefont
  {L.}~\bibnamefont {Cincio}}, \ and\ \bibinfo {author} {\bibfnamefont {P.~J.}\
  \bibnamefont {Coles}},\ }\href {https://arxiv.org/abs/2007.14384} {\bibfield
  {journal} {\bibinfo  {journal} {arXiv:2007.14384}\ } (\bibinfo {year}
  {2020}{\natexlab{a}})}\BibitemShut {NoStop}%
\bibitem [{\citenamefont {Sharma}\ \emph {et~al.}(2020)\citenamefont {Sharma},
  \citenamefont {Khatri}, \citenamefont {Cerezo},\ and\ \citenamefont
  {Coles}}]{sharma2020noise}%
  \BibitemOpen
  \bibfield  {author} {\bibinfo {author} {\bibfnamefont {K.}~\bibnamefont
  {Sharma}}, \bibinfo {author} {\bibfnamefont {S.}~\bibnamefont {Khatri}},
  \bibinfo {author} {\bibfnamefont {M.}~\bibnamefont {Cerezo}}, \ and\ \bibinfo
  {author} {\bibfnamefont {P.~J.}\ \bibnamefont {Coles}},\ }\href {\doibase
  10.1088/1367-2630/ab784c} {\bibfield  {journal} {\bibinfo  {journal} {New
  Journal of Physics}\ }\textbf {\bibinfo {volume} {22}},\ \bibinfo {pages}
  {043006} (\bibinfo {year} {2020})}\BibitemShut {NoStop}%
\bibitem [{\citenamefont {Gentini}\ \emph {et~al.}(2019)\citenamefont
  {Gentini}, \citenamefont {Cuccoli}, \citenamefont {Pirandola}, \citenamefont
  {Verrucchi},\ and\ \citenamefont {Banchi}}]{gentini2019noiseassisted}%
  \BibitemOpen
  \bibfield  {author} {\bibinfo {author} {\bibfnamefont {L.}~\bibnamefont
  {Gentini}}, \bibinfo {author} {\bibfnamefont {A.}~\bibnamefont {Cuccoli}},
  \bibinfo {author} {\bibfnamefont {S.}~\bibnamefont {Pirandola}}, \bibinfo
  {author} {\bibfnamefont {P.}~\bibnamefont {Verrucchi}}, \ and\ \bibinfo
  {author} {\bibfnamefont {L.}~\bibnamefont {Banchi}},\ }\href
  {https://arxiv.org/abs/1912.06744} {\bibfield  {journal} {\bibinfo  {journal}
  {arXiv:1912.06744}\ } (\bibinfo {year} {2019})}\BibitemShut {NoStop}%
\bibitem [{\citenamefont {Cerezo}\ \emph {et~al.}(2020)\citenamefont {Cerezo},
  \citenamefont {Sharma}, \citenamefont {Arrasmith},\ and\ \citenamefont
  {Coles}}]{cerezo2020variational}%
  \BibitemOpen
  \bibfield  {author} {\bibinfo {author} {\bibfnamefont {M.}~\bibnamefont
  {Cerezo}}, \bibinfo {author} {\bibfnamefont {K.}~\bibnamefont {Sharma}},
  \bibinfo {author} {\bibfnamefont {A.}~\bibnamefont {Arrasmith}}, \ and\
  \bibinfo {author} {\bibfnamefont {P.~J.}\ \bibnamefont {Coles}},\ }\href
  {https://arxiv.org/abs/2004.01372} {\bibfield  {journal} {\bibinfo  {journal}
  {arXiv:2004.01372}\ } (\bibinfo {year} {2020})}\BibitemShut {NoStop}%
\bibitem [{\citenamefont {Bravo-Prieto}\ \emph {et~al.}(2019)\citenamefont
  {Bravo-Prieto}, \citenamefont {LaRose}, \citenamefont {Cerezo}, \citenamefont
  {Subasi}, \citenamefont {Cincio},\ and\ \citenamefont
  {Coles}}]{bravo2019variational}%
  \BibitemOpen
  \bibfield  {author} {\bibinfo {author} {\bibfnamefont {C.}~\bibnamefont
  {Bravo-Prieto}}, \bibinfo {author} {\bibfnamefont {R.}~\bibnamefont
  {LaRose}}, \bibinfo {author} {\bibfnamefont {M.}~\bibnamefont {Cerezo}},
  \bibinfo {author} {\bibfnamefont {Y.}~\bibnamefont {Subasi}}, \bibinfo
  {author} {\bibfnamefont {L.}~\bibnamefont {Cincio}}, \ and\ \bibinfo {author}
  {\bibfnamefont {P.~J.}\ \bibnamefont {Coles}},\ }\href
  {https://arxiv.org/abs/1909.05820} {\bibfield  {journal} {\bibinfo  {journal}
  {arXiv:1909.05820}\ } (\bibinfo {year} {2019})}\BibitemShut {NoStop}%
\bibitem [{\citenamefont {Wang}\ \emph
  {et~al.}(2020{\natexlab{b}})\citenamefont {Wang}, \citenamefont {Song},\ and\
  \citenamefont {Wang}}]{wang2020variational}%
  \BibitemOpen
  \bibfield  {author} {\bibinfo {author} {\bibfnamefont {X.}~\bibnamefont
  {Wang}}, \bibinfo {author} {\bibfnamefont {Z.}~\bibnamefont {Song}}, \ and\
  \bibinfo {author} {\bibfnamefont {Y.}~\bibnamefont {Wang}},\ }\href
  {https://arxiv.org/abs/2006.02336} {\bibfield  {journal} {\bibinfo  {journal}
  {arXiv:2006.02336}\ } (\bibinfo {year} {2020}{\natexlab{b}})}\BibitemShut
  {NoStop}%
\bibitem [{\citenamefont {Anschuetz}\ \emph {et~al.}(2019)\citenamefont
  {Anschuetz}, \citenamefont {Olson}, \citenamefont {Aspuru-Guzik},\ and\
  \citenamefont {Cao}}]{anschuetz2019variational}%
  \BibitemOpen
  \bibfield  {author} {\bibinfo {author} {\bibfnamefont {E.}~\bibnamefont
  {Anschuetz}}, \bibinfo {author} {\bibfnamefont {J.}~\bibnamefont {Olson}},
  \bibinfo {author} {\bibfnamefont {A.}~\bibnamefont {Aspuru-Guzik}}, \ and\
  \bibinfo {author} {\bibfnamefont {Y.}~\bibnamefont {Cao}},\ }in\ \href@noop
  {} {\emph {\bibinfo {booktitle} {International Workshop on Quantum Technology
  and Optimization Problems}}}\ (\bibinfo {organization} {Springer},\ \bibinfo
  {year} {2019})\ pp.\ \bibinfo {pages} {74--85}\BibitemShut {NoStop}%
\bibitem [{\citenamefont {Zeng}\ \emph {et~al.}(2020)\citenamefont {Zeng},
  \citenamefont {Cao}, \citenamefont {Zhang}, \citenamefont {Xu},\ and\
  \citenamefont {Zeng}}]{zeng2020variational}%
  \BibitemOpen
  \bibfield  {author} {\bibinfo {author} {\bibfnamefont {J.}~\bibnamefont
  {Zeng}}, \bibinfo {author} {\bibfnamefont {C.}~\bibnamefont {Cao}}, \bibinfo
  {author} {\bibfnamefont {C.}~\bibnamefont {Zhang}}, \bibinfo {author}
  {\bibfnamefont {P.}~\bibnamefont {Xu}}, \ and\ \bibinfo {author}
  {\bibfnamefont {B.}~\bibnamefont {Zeng}},\ }\href
  {https://arxiv.org/abs/2008.09854} {\bibfield  {journal} {\bibinfo  {journal}
  {arXiv:2008.09854}\ } (\bibinfo {year} {2020})}\BibitemShut {NoStop}%
\bibitem [{\citenamefont {LaRose}\ \emph {et~al.}(2019)\citenamefont {LaRose},
  \citenamefont {Tikku}, \citenamefont {O'Neel-Judy}, \citenamefont {Cincio},\
  and\ \citenamefont {Coles}}]{larose2019variational}%
  \BibitemOpen
  \bibfield  {author} {\bibinfo {author} {\bibfnamefont {R.}~\bibnamefont
  {LaRose}}, \bibinfo {author} {\bibfnamefont {A.}~\bibnamefont {Tikku}},
  \bibinfo {author} {\bibfnamefont {{\'E}.}~\bibnamefont {O'Neel-Judy}},
  \bibinfo {author} {\bibfnamefont {L.}~\bibnamefont {Cincio}}, \ and\ \bibinfo
  {author} {\bibfnamefont {P.~J.}\ \bibnamefont {Coles}},\ }\href {\doibase
  10.1038/s41534-019-0167-6} {\bibfield  {journal} {\bibinfo  {journal} {npj
  Quantum Information}\ }\textbf {\bibinfo {volume} {5}},\ \bibinfo {pages}
  {57} (\bibinfo {year} {2019})}\BibitemShut {NoStop}%
\bibitem [{\citenamefont {Bravo-Prieto}\ \emph {et~al.}(2020)\citenamefont
  {Bravo-Prieto}, \citenamefont {Garcia-Martin},\ and\ \citenamefont
  {Latorre}}]{bravo2020quantum}%
  \BibitemOpen
  \bibfield  {author} {\bibinfo {author} {\bibfnamefont {C.}~\bibnamefont
  {Bravo-Prieto}}, \bibinfo {author} {\bibfnamefont {D.}~\bibnamefont
  {Garcia-Martin}}, \ and\ \bibinfo {author} {\bibfnamefont {J.}~\bibnamefont
  {Latorre}},\ }\href {https://link.aps.org/doi/10.1103/PhysRevA.101.062310}
  {\bibfield  {journal} {\bibinfo  {journal} {Phys. Rev. A}\ }\textbf {\bibinfo
  {volume} {101}},\ \bibinfo {pages} {062310} (\bibinfo {year}
  {2020})}\BibitemShut {NoStop}%
\bibitem [{\citenamefont {O'Malley}\ \emph {et~al.}(2016)\citenamefont
  {O'Malley}, \citenamefont {Babbush}, \citenamefont {Kivlichan}, \citenamefont
  {Romero}, \citenamefont {McClean}, \citenamefont {Barends}, \citenamefont
  {Kelly}, \citenamefont {Roushan}, \citenamefont {Tranter}, \citenamefont
  {Ding}, \citenamefont {Campbell}, \citenamefont {Chen}, \citenamefont {Chen},
  \citenamefont {Chiaro}, \citenamefont {Dunsworth}, \citenamefont {Fowler},
  \citenamefont {Jeffrey}, \citenamefont {Lucero}, \citenamefont {Megrant},
  \citenamefont {Mutus}, \citenamefont {Neeley}, \citenamefont {Neill},
  \citenamefont {Quintana}, \citenamefont {Sank}, \citenamefont {Vainsencher},
  \citenamefont {Wenner}, \citenamefont {White}, \citenamefont {Coveney},
  \citenamefont {Love}, \citenamefont {Neven}, \citenamefont {Aspuru-Guzik},\
  and\ \citenamefont {Martinis}}]{o2016scalable}%
  \BibitemOpen
  \bibfield  {author} {\bibinfo {author} {\bibfnamefont {P.~J.~J.}\
  \bibnamefont {O'Malley}}, \bibinfo {author} {\bibfnamefont {R.}~\bibnamefont
  {Babbush}}, \bibinfo {author} {\bibfnamefont {I.~D.}\ \bibnamefont
  {Kivlichan}}, \bibinfo {author} {\bibfnamefont {J.}~\bibnamefont {Romero}},
  \bibinfo {author} {\bibfnamefont {J.~R.}\ \bibnamefont {McClean}}, \bibinfo
  {author} {\bibfnamefont {R.}~\bibnamefont {Barends}}, \bibinfo {author}
  {\bibfnamefont {J.}~\bibnamefont {Kelly}}, \bibinfo {author} {\bibfnamefont
  {P.}~\bibnamefont {Roushan}}, \bibinfo {author} {\bibfnamefont
  {A.}~\bibnamefont {Tranter}}, \bibinfo {author} {\bibfnamefont
  {N.}~\bibnamefont {Ding}}, \bibinfo {author} {\bibfnamefont {B.}~\bibnamefont
  {Campbell}}, \bibinfo {author} {\bibfnamefont {Y.}~\bibnamefont {Chen}},
  \bibinfo {author} {\bibfnamefont {Z.}~\bibnamefont {Chen}}, \bibinfo {author}
  {\bibfnamefont {B.}~\bibnamefont {Chiaro}}, \bibinfo {author} {\bibfnamefont
  {A.}~\bibnamefont {Dunsworth}}, \bibinfo {author} {\bibfnamefont {A.~G.}\
  \bibnamefont {Fowler}}, \bibinfo {author} {\bibfnamefont {E.}~\bibnamefont
  {Jeffrey}}, \bibinfo {author} {\bibfnamefont {E.}~\bibnamefont {Lucero}},
  \bibinfo {author} {\bibfnamefont {A.}~\bibnamefont {Megrant}}, \bibinfo
  {author} {\bibfnamefont {J.~Y.}\ \bibnamefont {Mutus}}, \bibinfo {author}
  {\bibfnamefont {M.}~\bibnamefont {Neeley}}, \bibinfo {author} {\bibfnamefont
  {C.}~\bibnamefont {Neill}}, \bibinfo {author} {\bibfnamefont
  {C.}~\bibnamefont {Quintana}}, \bibinfo {author} {\bibfnamefont
  {D.}~\bibnamefont {Sank}}, \bibinfo {author} {\bibfnamefont {A.}~\bibnamefont
  {Vainsencher}}, \bibinfo {author} {\bibfnamefont {J.}~\bibnamefont {Wenner}},
  \bibinfo {author} {\bibfnamefont {T.~C.}\ \bibnamefont {White}}, \bibinfo
  {author} {\bibfnamefont {P.~V.}\ \bibnamefont {Coveney}}, \bibinfo {author}
  {\bibfnamefont {P.~J.}\ \bibnamefont {Love}}, \bibinfo {author}
  {\bibfnamefont {H.}~\bibnamefont {Neven}}, \bibinfo {author} {\bibfnamefont
  {A.}~\bibnamefont {Aspuru-Guzik}}, \ and\ \bibinfo {author} {\bibfnamefont
  {J.~M.}\ \bibnamefont {Martinis}},\ }\href {\doibase
  10.1103/PhysRevX.6.031007} {\bibfield  {journal} {\bibinfo  {journal} {Phys.
  Rev. X}\ }\textbf {\bibinfo {volume} {6}},\ \bibinfo {pages} {031007}
  (\bibinfo {year} {2016})}\BibitemShut {NoStop}%
\bibitem [{\citenamefont {Hempel}\ \emph {et~al.}(2018)\citenamefont {Hempel},
  \citenamefont {Maier}, \citenamefont {Romero}, \citenamefont {McClean},
  \citenamefont {Monz}, \citenamefont {Shen}, \citenamefont {Jurcevic},
  \citenamefont {Lanyon}, \citenamefont {Love}, \citenamefont {Babbush},
  \citenamefont {Aspuru-Guzik}, \citenamefont {Blatt},\ and\ \citenamefont
  {Roos}}]{Hempel2018Quantum}%
  \BibitemOpen
  \bibfield  {author} {\bibinfo {author} {\bibfnamefont {C.}~\bibnamefont
  {Hempel}}, \bibinfo {author} {\bibfnamefont {C.}~\bibnamefont {Maier}},
  \bibinfo {author} {\bibfnamefont {J.}~\bibnamefont {Romero}}, \bibinfo
  {author} {\bibfnamefont {J.}~\bibnamefont {McClean}}, \bibinfo {author}
  {\bibfnamefont {T.}~\bibnamefont {Monz}}, \bibinfo {author} {\bibfnamefont
  {H.}~\bibnamefont {Shen}}, \bibinfo {author} {\bibfnamefont {P.}~\bibnamefont
  {Jurcevic}}, \bibinfo {author} {\bibfnamefont {B.~P.}\ \bibnamefont
  {Lanyon}}, \bibinfo {author} {\bibfnamefont {P.}~\bibnamefont {Love}},
  \bibinfo {author} {\bibfnamefont {R.}~\bibnamefont {Babbush}}, \bibinfo
  {author} {\bibfnamefont {A.}~\bibnamefont {Aspuru-Guzik}}, \bibinfo {author}
  {\bibfnamefont {R.}~\bibnamefont {Blatt}}, \ and\ \bibinfo {author}
  {\bibfnamefont {C.~F.}\ \bibnamefont {Roos}},\ }\href {\doibase
  10.1103/PhysRevX.8.031022} {\bibfield  {journal} {\bibinfo  {journal} {Phys.
  Rev. X}\ }\textbf {\bibinfo {volume} {8}},\ \bibinfo {pages} {031022}
  (\bibinfo {year} {2018})}\BibitemShut {NoStop}%
\bibitem [{\citenamefont {Arute}\ \emph {et~al.}(2020)\citenamefont {Arute},
  \citenamefont {Arya}, \citenamefont {Babbush}, \citenamefont {Bacon},
  \citenamefont {Bardin}, \citenamefont {Barends}, \citenamefont {Boixo},
  \citenamefont {Broughton}, \citenamefont {Buckley}, \citenamefont {Buell}
  \emph {et~al.}}]{arute2020hartree}%
  \BibitemOpen
  \bibfield  {author} {\bibinfo {author} {\bibfnamefont {F.}~\bibnamefont
  {Arute}}, \bibinfo {author} {\bibfnamefont {K.}~\bibnamefont {Arya}},
  \bibinfo {author} {\bibfnamefont {R.}~\bibnamefont {Babbush}}, \bibinfo
  {author} {\bibfnamefont {D.}~\bibnamefont {Bacon}}, \bibinfo {author}
  {\bibfnamefont {J.~C.}\ \bibnamefont {Bardin}}, \bibinfo {author}
  {\bibfnamefont {R.}~\bibnamefont {Barends}}, \bibinfo {author} {\bibfnamefont
  {S.}~\bibnamefont {Boixo}}, \bibinfo {author} {\bibfnamefont
  {M.}~\bibnamefont {Broughton}}, \bibinfo {author} {\bibfnamefont {B.~B.}\
  \bibnamefont {Buckley}}, \bibinfo {author} {\bibfnamefont {D.~A.}\
  \bibnamefont {Buell}},  \emph {et~al.},\ }\href
  {https://arxiv.org/abs/2004.04174} {\bibfield  {journal} {\bibinfo  {journal}
  {arXiv:2004.04174}\ } (\bibinfo {year} {2020})}\BibitemShut {NoStop}%
\bibitem [{\citenamefont {Bremner}\ \emph {et~al.}(2011)\citenamefont
  {Bremner}, \citenamefont {Jozsa},\ and\ \citenamefont
  {Shepherd}}]{Bremner2011Classical}%
  \BibitemOpen
  \bibfield  {author} {\bibinfo {author} {\bibfnamefont {M.~J.}\ \bibnamefont
  {Bremner}}, \bibinfo {author} {\bibfnamefont {R.}~\bibnamefont {Jozsa}}, \
  and\ \bibinfo {author} {\bibfnamefont {D.~J.}\ \bibnamefont {Shepherd}},\
  }\href {\doibase 10.1098/rspa.2010.0301} {\bibfield  {journal} {\bibinfo
  {journal} {Proceedings of the Royal Society A: Mathematical, Physical and
  Engineering Sciences}\ }\textbf {\bibinfo {volume} {467}},\ \bibinfo {pages}
  {459} (\bibinfo {year} {2011})}\BibitemShut {NoStop}%
\bibitem [{\citenamefont {Aaronson}\ and\ \citenamefont
  {Arkhipov}(2011)}]{aaronson2011computational}%
  \BibitemOpen
  \bibfield  {author} {\bibinfo {author} {\bibfnamefont {S.}~\bibnamefont
  {Aaronson}}\ and\ \bibinfo {author} {\bibfnamefont {A.}~\bibnamefont
  {Arkhipov}},\ }in\ \href@noop {} {\emph {\bibinfo {booktitle} {Proceedings of
  the forty-third annual ACM symposium on Theory of computing}}}\ (\bibinfo
  {year} {2011})\ pp.\ \bibinfo {pages} {333--342}\BibitemShut {NoStop}%
\bibitem [{\citenamefont {Fujii}\ and\ \citenamefont
  {Morimae}(2017)}]{Fujii2017Commuting}%
  \BibitemOpen
  \bibfield  {author} {\bibinfo {author} {\bibfnamefont {K.}~\bibnamefont
  {Fujii}}\ and\ \bibinfo {author} {\bibfnamefont {T.}~\bibnamefont
  {Morimae}},\ }\href {\doibase 10.1088/1367-2630/aa5fdb} {\bibfield  {journal}
  {\bibinfo  {journal} {New Journal of Physics}\ }\textbf {\bibinfo {volume}
  {19}},\ \bibinfo {pages} {033003} (\bibinfo {year} {2017})}\BibitemShut
  {NoStop}%
\bibitem [{\citenamefont {Bremner}\ \emph {et~al.}(2016)\citenamefont
  {Bremner}, \citenamefont {Montanaro},\ and\ \citenamefont
  {Shepherd}}]{Bremner2016Average}%
  \BibitemOpen
  \bibfield  {author} {\bibinfo {author} {\bibfnamefont {M.~J.}\ \bibnamefont
  {Bremner}}, \bibinfo {author} {\bibfnamefont {A.}~\bibnamefont {Montanaro}},
  \ and\ \bibinfo {author} {\bibfnamefont {D.~J.}\ \bibnamefont {Shepherd}},\
  }\href {\doibase 10.1103/PhysRevLett.117.080501} {\bibfield  {journal}
  {\bibinfo  {journal} {Phys. Rev. Lett.}\ }\textbf {\bibinfo {volume} {117}},\
  \bibinfo {pages} {080501} (\bibinfo {year} {2016})}\BibitemShut {NoStop}%
\bibitem [{\citenamefont {Aaronson}\ and\ \citenamefont
  {Chen}(2016)}]{aaronson2016complexitytheoretic}%
  \BibitemOpen
  \bibfield  {author} {\bibinfo {author} {\bibfnamefont {S.}~\bibnamefont
  {Aaronson}}\ and\ \bibinfo {author} {\bibfnamefont {L.}~\bibnamefont
  {Chen}},\ }\href {https://arxiv.org/abs/1612.05903} {\bibfield  {journal}
  {\bibinfo  {journal} {arxiv:1612.05903}\ } (\bibinfo {year}
  {2016})}\BibitemShut {NoStop}%
\bibitem [{\citenamefont {Kandala}\ \emph {et~al.}(2019)\citenamefont
  {Kandala}, \citenamefont {Temme}, \citenamefont {C{\'o}rcoles}, \citenamefont
  {Mezzacapo}, \citenamefont {Chow},\ and\ \citenamefont
  {Gambetta}}]{kandala2019error}%
  \BibitemOpen
  \bibfield  {author} {\bibinfo {author} {\bibfnamefont {A.}~\bibnamefont
  {Kandala}}, \bibinfo {author} {\bibfnamefont {K.}~\bibnamefont {Temme}},
  \bibinfo {author} {\bibfnamefont {A.~D.}\ \bibnamefont {C{\'o}rcoles}},
  \bibinfo {author} {\bibfnamefont {A.}~\bibnamefont {Mezzacapo}}, \bibinfo
  {author} {\bibfnamefont {J.~M.}\ \bibnamefont {Chow}}, \ and\ \bibinfo
  {author} {\bibfnamefont {J.~M.}\ \bibnamefont {Gambetta}},\ }\href
  {https://doi.org/10.1038/s41586-019-1040-7} {\bibfield  {journal} {\bibinfo
  {journal} {Nature}\ }\textbf {\bibinfo {volume} {567}},\ \bibinfo {pages}
  {491} (\bibinfo {year} {2019})}\BibitemShut {NoStop}%
\bibitem [{\citenamefont {Temme}\ \emph {et~al.}(2017)\citenamefont {Temme},
  \citenamefont {Bravyi},\ and\ \citenamefont {Gambetta}}]{Temme2017Error}%
  \BibitemOpen
  \bibfield  {author} {\bibinfo {author} {\bibfnamefont {K.}~\bibnamefont
  {Temme}}, \bibinfo {author} {\bibfnamefont {S.}~\bibnamefont {Bravyi}}, \
  and\ \bibinfo {author} {\bibfnamefont {J.~M.}\ \bibnamefont {Gambetta}},\
  }\href {\doibase 10.1103/PhysRevLett.119.180509} {\bibfield  {journal}
  {\bibinfo  {journal} {Phys. Rev. Lett.}\ }\textbf {\bibinfo {volume} {119}},\
  \bibinfo {pages} {180509} (\bibinfo {year} {2017})}\BibitemShut {NoStop}%
\bibitem [{\citenamefont {Endo}\ \emph {et~al.}(2018)\citenamefont {Endo},
  \citenamefont {Benjamin},\ and\ \citenamefont {Li}}]{Endo2018Practical}%
  \BibitemOpen
  \bibfield  {author} {\bibinfo {author} {\bibfnamefont {S.}~\bibnamefont
  {Endo}}, \bibinfo {author} {\bibfnamefont {S.~C.}\ \bibnamefont {Benjamin}},
  \ and\ \bibinfo {author} {\bibfnamefont {Y.}~\bibnamefont {Li}},\ }\href
  {\doibase 10.1103/PhysRevX.8.031027} {\bibfield  {journal} {\bibinfo
  {journal} {Phys. Rev. X}\ }\textbf {\bibinfo {volume} {8}},\ \bibinfo {pages}
  {031027} (\bibinfo {year} {2018})}\BibitemShut {NoStop}%
\bibitem [{\citenamefont {Schuld}\ \emph {et~al.}(2019)\citenamefont {Schuld},
  \citenamefont {Bergholm}, \citenamefont {Gogolin}, \citenamefont {Izaac},\
  and\ \citenamefont {Killoran}}]{Schuld2019}%
  \BibitemOpen
  \bibfield  {author} {\bibinfo {author} {\bibfnamefont {M.}~\bibnamefont
  {Schuld}}, \bibinfo {author} {\bibfnamefont {V.}~\bibnamefont {Bergholm}},
  \bibinfo {author} {\bibfnamefont {C.}~\bibnamefont {Gogolin}}, \bibinfo
  {author} {\bibfnamefont {J.}~\bibnamefont {Izaac}}, \ and\ \bibinfo {author}
  {\bibfnamefont {N.}~\bibnamefont {Killoran}},\ }\href {\doibase
  10.1103/PhysRevA.99.032331} {\bibfield  {journal} {\bibinfo  {journal} {Phys.
  Rev. A}\ }\textbf {\bibinfo {volume} {99}},\ \bibinfo {pages} {032331}
  (\bibinfo {year} {2019})}\BibitemShut {NoStop}%
\bibitem [{\citenamefont {Mitarai}\ \emph {et~al.}(2018)\citenamefont
  {Mitarai}, \citenamefont {Negoro}, \citenamefont {Kitagawa},\ and\
  \citenamefont {Fujii}}]{Mitarai2018}%
  \BibitemOpen
  \bibfield  {author} {\bibinfo {author} {\bibfnamefont {K.}~\bibnamefont
  {Mitarai}}, \bibinfo {author} {\bibfnamefont {M.}~\bibnamefont {Negoro}},
  \bibinfo {author} {\bibfnamefont {M.}~\bibnamefont {Kitagawa}}, \ and\
  \bibinfo {author} {\bibfnamefont {K.}~\bibnamefont {Fujii}},\ }\href
  {\doibase 10.1103/PhysRevA.98.032309} {\bibfield  {journal} {\bibinfo
  {journal} {Phys. Rev. A}\ }\textbf {\bibinfo {volume} {98}},\ \bibinfo
  {pages} {032309} (\bibinfo {year} {2018})}\BibitemShut {NoStop}%
\bibitem [{\citenamefont {{Kingma}}\ and\ \citenamefont
  {{Ba}}(2014)}]{Kingma2014Adam}%
  \BibitemOpen
  \bibfield  {author} {\bibinfo {author} {\bibfnamefont {D.~P.}\ \bibnamefont
  {{Kingma}}}\ and\ \bibinfo {author} {\bibfnamefont {J.}~\bibnamefont
  {{Ba}}},\ }\href {https://arxiv.org/abs/1412.6980} {\bibfield  {journal}
  {\bibinfo  {journal} {arXiv:1412.6980}\ } (\bibinfo {year}
  {2014})}\BibitemShut {NoStop}%
\bibitem [{\citenamefont {Cao}\ \emph {et~al.}(2019)\citenamefont {Cao},
  \citenamefont {Romero}, \citenamefont {Olson}, \citenamefont {Degroote},
  \citenamefont {Johnson}, \citenamefont {Kieferová}, \citenamefont
  {Kivlichan}, \citenamefont {Menke}, \citenamefont {Peropadre}, \citenamefont
  {Sawaya}, \citenamefont {Sim}, \citenamefont {Veis},\ and\ \citenamefont
  {Aspuru-Guzik}}]{Cao2019Quantum}%
  \BibitemOpen
  \bibfield  {author} {\bibinfo {author} {\bibfnamefont {Y.}~\bibnamefont
  {Cao}}, \bibinfo {author} {\bibfnamefont {J.}~\bibnamefont {Romero}},
  \bibinfo {author} {\bibfnamefont {J.~P.}\ \bibnamefont {Olson}}, \bibinfo
  {author} {\bibfnamefont {M.}~\bibnamefont {Degroote}}, \bibinfo {author}
  {\bibfnamefont {P.~D.}\ \bibnamefont {Johnson}}, \bibinfo {author}
  {\bibfnamefont {M.}~\bibnamefont {Kieferová}}, \bibinfo {author}
  {\bibfnamefont {I.~D.}\ \bibnamefont {Kivlichan}}, \bibinfo {author}
  {\bibfnamefont {T.}~\bibnamefont {Menke}}, \bibinfo {author} {\bibfnamefont
  {B.}~\bibnamefont {Peropadre}}, \bibinfo {author} {\bibfnamefont {N.~P.~D.}\
  \bibnamefont {Sawaya}}, \bibinfo {author} {\bibfnamefont {S.}~\bibnamefont
  {Sim}}, \bibinfo {author} {\bibfnamefont {L.}~\bibnamefont {Veis}}, \ and\
  \bibinfo {author} {\bibfnamefont {A.}~\bibnamefont {Aspuru-Guzik}},\ }\href
  {\doibase 10.1021/acs.chemrev.8b00803} {\bibfield  {journal} {\bibinfo
  {journal} {Chemical Reviews}\ }\textbf {\bibinfo {volume} {119}},\ \bibinfo
  {pages} {10856} (\bibinfo {year} {2019})}\BibitemShut {NoStop}%
\bibitem [{\citenamefont {Sim}\ \emph {et~al.}(2019)\citenamefont {Sim},
  \citenamefont {Johnson},\ and\ \citenamefont
  {Aspuru-Guzik}}]{Sim2019Expressibility}%
  \BibitemOpen
  \bibfield  {author} {\bibinfo {author} {\bibfnamefont {S.}~\bibnamefont
  {Sim}}, \bibinfo {author} {\bibfnamefont {P.~D.}\ \bibnamefont {Johnson}}, \
  and\ \bibinfo {author} {\bibfnamefont {A.}~\bibnamefont {Aspuru-Guzik}},\
  }\href {\doibase 10.1002/qute.201900070} {\bibfield  {journal} {\bibinfo
  {journal} {Advanced Quantum Technologies}\ }\textbf {\bibinfo {volume} {2}},\
  \bibinfo {pages} {1900070} (\bibinfo {year} {2019})}\BibitemShut {NoStop}%
\bibitem [{\citenamefont {Romero}\ \emph {et~al.}(2018)\citenamefont {Romero},
  \citenamefont {Babbush}, \citenamefont {McClean}, \citenamefont {Hempel},
  \citenamefont {Love},\ and\ \citenamefont
  {Aspuru-Guzik}}]{Romero2018Strategies}%
  \BibitemOpen
  \bibfield  {author} {\bibinfo {author} {\bibfnamefont {J.}~\bibnamefont
  {Romero}}, \bibinfo {author} {\bibfnamefont {R.}~\bibnamefont {Babbush}},
  \bibinfo {author} {\bibfnamefont {J.~R.}\ \bibnamefont {McClean}}, \bibinfo
  {author} {\bibfnamefont {C.}~\bibnamefont {Hempel}}, \bibinfo {author}
  {\bibfnamefont {P.~J.}\ \bibnamefont {Love}}, \ and\ \bibinfo {author}
  {\bibfnamefont {A.}~\bibnamefont {Aspuru-Guzik}},\ }\href {\doibase
  10.1088/2058-9565/aad3e4} {\bibfield  {journal} {\bibinfo  {journal} {Quantum
  Science and Technology}\ }\textbf {\bibinfo {volume} {4}},\ \bibinfo {pages}
  {014008} (\bibinfo {year} {2018})}\BibitemShut {NoStop}%
\bibitem [{\citenamefont {Wecker}\ \emph {et~al.}(2015)\citenamefont {Wecker},
  \citenamefont {Hastings},\ and\ \citenamefont {Troyer}}]{Wecker2015Progress}%
  \BibitemOpen
  \bibfield  {author} {\bibinfo {author} {\bibfnamefont {D.}~\bibnamefont
  {Wecker}}, \bibinfo {author} {\bibfnamefont {M.~B.}\ \bibnamefont
  {Hastings}}, \ and\ \bibinfo {author} {\bibfnamefont {M.}~\bibnamefont
  {Troyer}},\ }\href {\doibase 10.1103/PhysRevA.92.042303} {\bibfield
  {journal} {\bibinfo  {journal} {Phys. Rev. A}\ }\textbf {\bibinfo {volume}
  {92}},\ \bibinfo {pages} {042303} (\bibinfo {year} {2015})}\BibitemShut
  {NoStop}%
\bibitem [{\citenamefont {Dallaire-Demers}\ \emph {et~al.}(2019)\citenamefont
  {Dallaire-Demers}, \citenamefont {Romero}, \citenamefont {Veis},
  \citenamefont {Sim},\ and\ \citenamefont {Aspuru-Guzik}}]{Dallaire2019LDCA}%
  \BibitemOpen
  \bibfield  {author} {\bibinfo {author} {\bibfnamefont {P.-L.}\ \bibnamefont
  {Dallaire-Demers}}, \bibinfo {author} {\bibfnamefont {J.}~\bibnamefont
  {Romero}}, \bibinfo {author} {\bibfnamefont {L.}~\bibnamefont {Veis}},
  \bibinfo {author} {\bibfnamefont {S.}~\bibnamefont {Sim}}, \ and\ \bibinfo
  {author} {\bibfnamefont {A.}~\bibnamefont {Aspuru-Guzik}},\ }\href {\doibase
  10.1088/2058-9565/ab3951} {\bibfield  {journal} {\bibinfo  {journal} {Quantum
  Science and Technology}\ }\textbf {\bibinfo {volume} {4}},\ \bibinfo {pages}
  {045005} (\bibinfo {year} {2019})}\BibitemShut {NoStop}%
\bibitem [{\citenamefont {Grimsley}\ \emph {et~al.}(2019)\citenamefont
  {Grimsley}, \citenamefont {Economou}, \citenamefont {Barnes},\ and\
  \citenamefont {Mayhall}}]{Grimsley2019Adaptive}%
  \BibitemOpen
  \bibfield  {author} {\bibinfo {author} {\bibfnamefont {H.~R.}\ \bibnamefont
  {Grimsley}}, \bibinfo {author} {\bibfnamefont {S.~E.}\ \bibnamefont
  {Economou}}, \bibinfo {author} {\bibfnamefont {E.}~\bibnamefont {Barnes}}, \
  and\ \bibinfo {author} {\bibfnamefont {N.~J.}\ \bibnamefont {Mayhall}},\
  }\href {https://doi.org/10.1038/s41467-019-10988-2} {\bibfield  {journal}
  {\bibinfo  {journal} {Nature Communications}\ }\textbf {\bibinfo {volume}
  {10}},\ \bibinfo {pages} {3007} (\bibinfo {year} {2019})}\BibitemShut
  {NoStop}%
\bibitem [{\citenamefont {Tang}\ \emph {et~al.}(2020)\citenamefont {Tang},
  \citenamefont {Shkolnikov}, \citenamefont {Barron}, \citenamefont {Grimsley},
  \citenamefont {Mayhall}, \citenamefont {Barnes},\ and\ \citenamefont
  {Economou}}]{tang2020qubitadaptvqe}%
  \BibitemOpen
  \bibfield  {author} {\bibinfo {author} {\bibfnamefont {H.~L.}\ \bibnamefont
  {Tang}}, \bibinfo {author} {\bibfnamefont {V.~O.}\ \bibnamefont
  {Shkolnikov}}, \bibinfo {author} {\bibfnamefont {G.~S.}\ \bibnamefont
  {Barron}}, \bibinfo {author} {\bibfnamefont {H.~R.}\ \bibnamefont
  {Grimsley}}, \bibinfo {author} {\bibfnamefont {N.~J.}\ \bibnamefont
  {Mayhall}}, \bibinfo {author} {\bibfnamefont {E.}~\bibnamefont {Barnes}}, \
  and\ \bibinfo {author} {\bibfnamefont {S.~E.}\ \bibnamefont {Economou}},\
  }\href {https://arxiv.org/abs/1911.10205} {\bibfield  {journal} {\bibinfo
  {journal} {arXiv:1911.10205}\ } (\bibinfo {year} {2020})}\BibitemShut
  {NoStop}%
\bibitem [{\citenamefont {Nielsen}\ and\ \citenamefont
  {Chuang}(2010)}]{nielsen_chuang_2010}%
  \BibitemOpen
  \bibfield  {author} {\bibinfo {author} {\bibfnamefont {M.~A.}\ \bibnamefont
  {Nielsen}}\ and\ \bibinfo {author} {\bibfnamefont {I.~L.}\ \bibnamefont
  {Chuang}},\ }\href {\doibase 10.1017/CBO9780511976667} {\emph {\bibinfo
  {title} {Quantum Computation and Quantum Information: 10th Anniversary
  Edition}}}\ (\bibinfo  {publisher} {Cambridge University Press},\ \bibinfo
  {year} {2010})\BibitemShut {NoStop}%
\bibitem [{\citenamefont {{IBM Q Experience}}()}]{ibmq}%
  \BibitemOpen
  \bibfield  {author} {\bibinfo {author} {\bibnamefont {{IBM Q Experience}}},\
  }\href {https://quantum-computing.ibm.com} {\bibinfo  {journal}
  {https://quantum-computing.ibm.com}\ }\BibitemShut {NoStop}%
\bibitem [{\citenamefont {{Ayral}}\ \emph {et~al.}(2020)\citenamefont
  {{Ayral}}, \citenamefont {{Le Régent}}, \citenamefont {{Saleem}},
  \citenamefont {{Alexeev}},\ and\ \citenamefont
  {{Suchara}}}]{Ayral2020Quantum}%
  \BibitemOpen
\bibfield  {journal} {  }\bibfield  {author} {\bibinfo {author} {\bibfnamefont
  {T.}~\bibnamefont {{Ayral}}}, \bibinfo {author} {\bibfnamefont {F.~M.}\
  \bibnamefont {{Le Régent}}}, \bibinfo {author} {\bibfnamefont
  {Z.}~\bibnamefont {{Saleem}}}, \bibinfo {author} {\bibfnamefont
  {Y.}~\bibnamefont {{Alexeev}}}, \ and\ \bibinfo {author} {\bibfnamefont
  {M.}~\bibnamefont {{Suchara}}},\ }\bibfield  {booktitle} {\emph {\bibinfo
  {booktitle} {2020 IEEE Computer Society Annual Symposium on VLSI (ISVLSI)}},\
  }\href {\doibase 10.1109/ISVLSI49217.2020.00034} {\ ,\ \bibinfo {pages} {138}
  (\bibinfo {year} {2020})}\BibitemShut {NoStop}%
\end{thebibliography}%

\end{document}